\documentclass[sigconf]{acmart}

\usepackage{booktabs} 
\usepackage{lipsum}
\usepackage{listings}
\usepackage{float}
\usepackage{url}
\usepackage{todonotes}
\usepackage{multicol}
\usepackage{color,soul}
\usepackage{amssymb,amsmath}

\newcommand{\todoi}[1]{\todo[inline]{#1}}
\newcommand{\ignore}[1]{}

\setstcolor{orange}
\sethlcolor{yellow}
\usepackage{subcaption}

\usepackage{colortbl} 
\usepackage{fancyvrb}

\definecolor{grayborder}{rgb}{0.75,0.75,0.75}
\DefineVerbatimEnvironment
{verb}{Verbatim}
{fontsize=\scriptsize,tabsize=4,frame=single,framesep=1mm,rulecolor=\color{grayborder},baselinestretch=1,commandchars=\\\[\]}
\DefineVerbatimEnvironment
{verb2}{Verbatim}
{fontsize=\scriptsize,tabsize=4,frame=single,framesep=1mm,rulecolor=\color{grayborder},baselinestretch=1,commandchars=\\<>}
\DefineVerbatimEnvironment
{verbnocommand}{Verbatim}
{fontsize=\scriptsize,tabsize=4,frame=single,framesep=1mm,rulecolor=\color{grayborder},baselinestretch=1}

\usepackage{relsize}
\usepackage{xspace}
\newcommand{\Rplus}{\protect\hspace{-.1em}\protect\raisebox{.35ex}{\smaller{\smaller\textbf{+}}}}
\newcommand{\Cpp}{\mbox{C\Rplus\Rplus}\xspace}

\makeatletter
\let\old@citex\@citex
\makeatother
\usepackage[english]{babel}
\makeatletter
\let\@citex\old@citex
\makeatother

\def\Skip{\par\smallskip\nobreak\noindent}
\def\paragraph{\Skip\textbf}
\def\citep{\cite}

\renewcommand\footnotetextcopyrightpermission[1]{}
\acmConference[]{}
\acmYear{}
\begin{abstract}

We present a novel architecture, \emph{In-Database Entity
 Linking (IDEL)}, in which we  integrate the analytics-optimized RDBMS MonetDB with
 neural text mining abilities. 
Our system design abstracts core tasks of most neural entity linking systems for MonetDB. To the best of our knowledge, this is the first defacto implemented system integrating entity-linking in a database. We leverage the ability of MonetDB to support in-database-analytics with user defined functions (UDFs) implemented in Python. These functions call machine learning libraries for neural text mining, such as TensorFlow.
The system achieves zero cost for data shipping and transformation by utilizing
 MonetDB's ability to embed Python processes in the database kernel and
 exchange data in NumPy arrays.
IDEL represents text and relational data in a joint vector space with neural
 embeddings and can compensate errors with ambiguous entity representations.
For detecting matching entities, we propose a novel similarity function based
 on joint neural embeddings which are learned via minimizing pairwise
 contrastive ranking loss. This function utilizes a high dimensional index structures for fast retrieval of matching entities.
Our first implementation and experiments using the WebNLG corpus show the
 effectiveness and the potentials of IDEL.

\end{abstract}

\begin{document}
\title{IDEL: In-Database Entity Linking with Neural Embeddings}
\titlenote{This manuscript is a preprint for a paper submitted to VLDB2018}

\author{Torsten Kilias, Alexander L\"oser, Felix A. Gers}
\affiliation{%
  \institution{Beuth University of Applied Sciences}
  \city{Luxemburger Stra{\ss}e 10, Berlin}
  \postcode{13353}
}
\email{[tkilias,aloeser,gers]@beuth-hochschule.de}

\author{R. Koopmanschap, Y. Zhang, M. Kersten}
\affiliation{%
  \institution{MonetDB Solutions}
  \city{Science Park 123, Amsterdam}
  \postcode{1098XG}
}
\email{[firstletter.lastName]@monetdbsolutions.com}

\renewcommand{\shortauthors}{T. Kilias et al.}

\date{28 February 2018}
\maketitle

\section{Introduction}
\label{sec:introduction}

A particular exciting source for complementing relational data is text data.
There are endless opportunities for enterprises in various domains to gain
 advantages for their offerings or operations, when the enterprise relational
 data can be linked to the abundance of text data on the web.
The entities represented in the relational data can then be complemented,
 updated and extended with the information in the text data (which is often
 more fresh).

\emph{Brand- and competitor monitoring.}
Fashion retailers must closely observe news on the Web about their own brands and
 products, as well as those of their competitors.
Typically, information of the brands and products are kept in some relational
 tables (e.g. product catalogs).
Next to that, a fashion retailer can retrieve relevant texts (e.g. news
 articles, blogs and customer reviews about fashion items) from the Web.
With some text analysis tools, the fashion retailer would be able to enrich her
 product catalog with information from the text data to improve search
 results.
In addition, by monitoring the public opinions about fashion items, the fashion
 retailer would be able to adjust her products to the trends more timely and
 accurately.

\emph{Disaster monitoring.}
For reinsurance companies (i.e. companies insuring other insurance companies),
 anticipating potentially huge claims from their insurance customers after some
 (e.g. natural) disasters is a core business.
Next to collecting extensive historical data, such as customer information and
 various statistics, reinsurance companies also closely monitor news messages as
 an additional data source to predict probabilities of losses.
For instance, when a container ship has had an accident, a reinsurance company
 would want to know if this will affect her customers. Analogously, fires burning down entire buildings, such as important warehouses, cause relevant losses for reinsurance companies. These events can break a supply chain, therefore reinsurance company will most probably
 receive claims from her affected customers for their losses.  Reinsurance companies want to map text reports about companies and
 disasters to their databases, so that their analysts can search for
 information in both text and tabular data.
In addition, those analysts want to combine fresh information from local
 newspaper about companies in their database.
So news messages serve here as a source of fairly fresh data, in contrast to
 statistics/financial reports written quarterly or even annually.

\begin{figure}
	\center\includegraphics[width=\columnwidth]{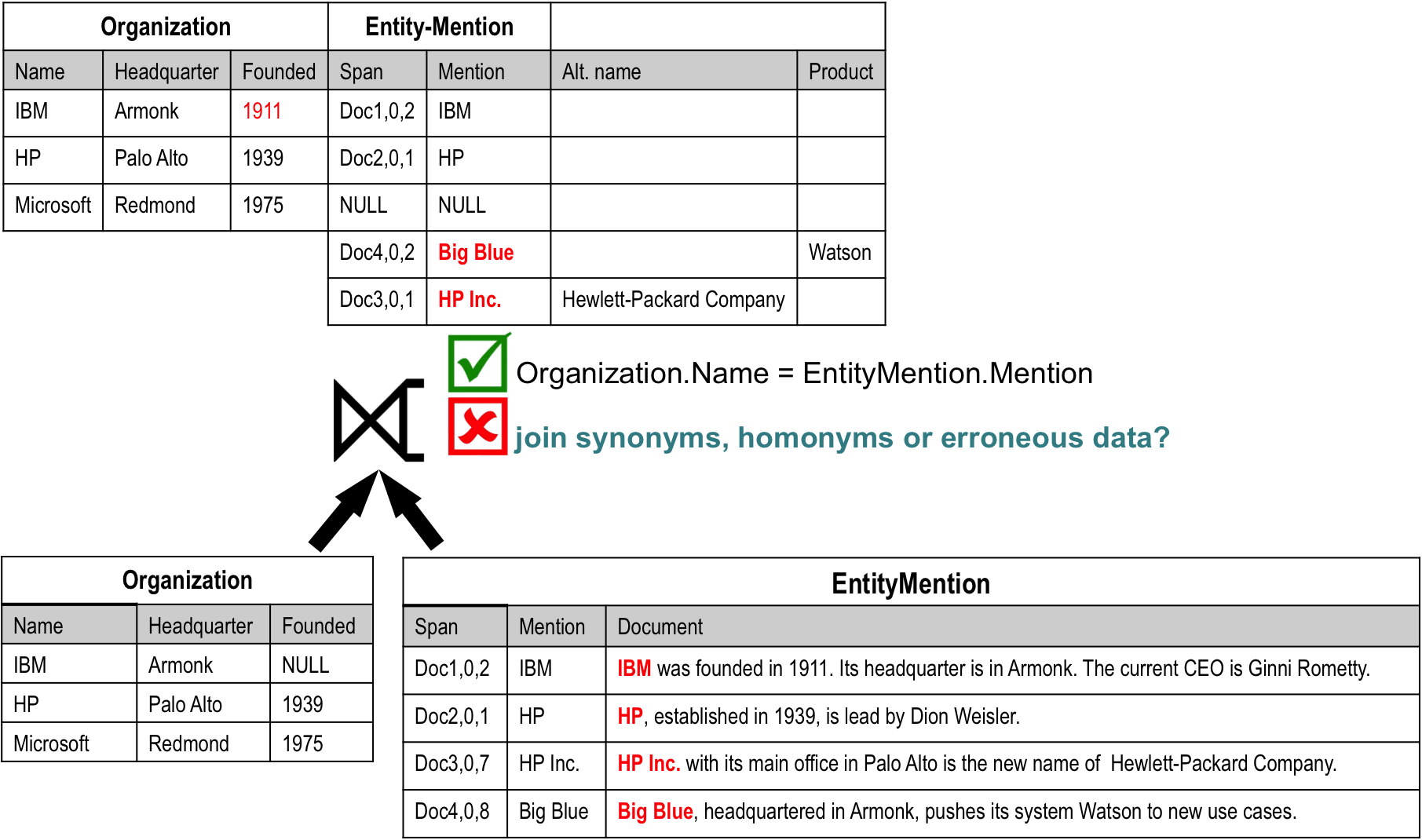}
	\caption{An example of joining relational data (i.e. {\sf Organization}) with
	text data (i.e. {\sf EntityMention}). An exact match strategy would have low recall, since it would not be able to match `Big Blue' with `IBM' or `HP Inc.' with `HP'.}
	\label{fig:textjoinexample}
\end{figure}

\Skip
\paragraph{Entity linking between text and tables} To realize these kinds of applications, a basic step is linking entities
 mentioned in text data to entities represented in relational data, so that
 missing data in the relational tables can be filled in or new data can be
 added.
An example is shown in Figure~\ref{fig:textjoinexample}, where the text data
 (i.e. {\sf Document}) is already preprocessed by some entity
 recognizers \cite{DBLP:conf/coling/ArnoldDL16}, which annotate the text data with entities
 recognized (i.e. {\sf Mention}) and their positions in the text (i.e. {\sf
 Span}).
However, further process of linking the recognized entities to the relational
 entities requires advanced text join techniques.
The main challenge is to compute fuzzy joins from synonyms, homonyms or even
 erroneous texts.

\emph{Classic text join.}
An often practiced but limited solution to this problem is classic (i.e.
 linguistic features based) text join techniques~\cite{kilias2015indrex},
 \cite{chiticariu2010systemt}.
First, each entity in the relational data is represented as a character
 sequence; while a text join system is used to extract candidate \emph{entity
 mentions} from the text data.
Next, the text join system executes a cross join between the two sets of
 entities.
The step often produces a large set of matching candidates.
So, finally, users can apply some lexical filter condition, such as an exact or
 a containment match, to reduce the result size.
However, this practice often suffers from low recall and precision when faced
 with ambiguous entity type families and entity mentions.
Typical error sources are, for instance, matches between
 \emph{homonyms} (e.g. IBM as the IATA airport code or the IT company),
 \emph{hyponyms} (e.g. ``SAP SE'' vs. ``SAP Deutschland SE \& Co. KG''),
 \emph{synonyms} (e.g. IBM and ``Big Blue'') and
 \emph{misspellings in text} (e.g. ``SAP Dtld. SE \& Co. KG'').

\emph{Entity-linking (EL)} between text and a more structured representation
 has been an active research topic for many years in the web community and
 among Computational Linguists \cite{DBLP:conf/sigmod/RenEJH16}.
EL systems abstract the entity linking problem as a multi-value multi-class
 classification task \cite{DBLP:conf/emnlp/GuptaSR17}.
Given a relational and a textual representation of entities, an EL system
 learns a similarity model for a classifier that can determine if there is a
 match or not.

\Skip
 \textbf{The need for Neural Entity Linking in RDBMSs}
However, existing EL systems come stand-alone or as separate tools, while so
 far, there has been little support inside RDBMSs for such advanced natural
 language processing (NLP) tasks. First, users (e.g. data scientists) often \emph{have to use three systems}: one
 for relational data (RDBMS), one for text data (often Lucene) and one for EL tasks (often homegrown). The first problem comes with the choice of a proper EL-system. Although there are many research papers and a few prototypes for several domains available, most work comes with domain specific features, ontologies or dictionaries, and is often not directly applicable for linking data from a particular database or needs extensive  fine tuning for a particular domain. First, to apply EL on relational data, a user has to move the data from the RDBMS to the EL tool. Not only, this will \emph{require significant human and technical resources for shipping data}. Worse, \emph{very few data scientists receive a proper training in both worlds}. Finding data scientists with proper training in entity linking and with deep knowledge in RDBMS is difficult. Moreover, \emph{transforming textual data in a relational representation requires glue and development time} from data scientists to bind these different system landscapes seamlessly. Finally, domain specific information extraction is an iterative task. It requires to continuously \emph{adopt manually tweaked features for recognizing entities}. A result from the above discussion is that many projects that combine textual data with existing relational data may
likely fail and/or be infeasible.

Overall, current approaches to EL have many technical drawbacks, e.g. high maintenance, data
 provenance problems, bad scalability due to data conversion, transferring and
 storage costs, and also non-technical drawbacks, such as the difficulty of hiring people trained in both the RDBMS and the NLP worlds.

\Skip
\textbf{Our contributions }
Ideally, the system would execute EL without any domain specific feature engineering, only triggered by a SQL query and in a single system without costly data shipping.
In this paper, we propose \emph{in-database entity linking (IDEL)}, a single
 system in which relational data, text data and entity linking tools are
 integrated.  IDEL stores both relational and text data in MonetDB, an open-source RDBMS
 optimized for in-memory analytics.
Entity linking components are tightly integrated into the kernel of MonetDB
 through SQL user defined functions (UDFs) implemented in
 Python~\cite{Raasveldt:2016:VUC:2949689.2949703}.
In this way, various neural machine learning libraries, e.g. TensorFlow,
 can be used to facilitate entity linking with neural embeddings. We chose neural embeddings, since the system will learn `features' from existing signals in relational and text data as hidden layers in a neural network and therefore can reduce human costs for feature engineering drastically.

In IDEL, we choose the RDBMS as the basis of the architecture, and integrating
 text data and entity linking into it for several carefully considered reasons.
First, while IDEL is a generally applicable architecture for many text analysis
 applications, we primarily target at enterprise applications, in which
 enterprises mainly want to enrich their fairly static relational data with
 information from dynamic text data.
Since traditionally enterprise data is already stored in an RDBMS, an RDBMS
 based architecture has the biggest potential of a seamless adaptation.
Second, an RDBMS has an extensive and powerful engine for pre- and
 post-entity-linking query processing and data analysis.
Finally, in-database analytics (i.e. bring the computation as close as possible
 to the data instead of moving the data to the computation) has long been
 recognized as the way-to-go for big data analytics.
Following the same philosophy, we propose an in-database entity linking
 architecture, which can directly benefit from existing in-database analytics
 features.

\begin{figure*}[t!]
	\centering\includegraphics[width=0.8\textwidth]{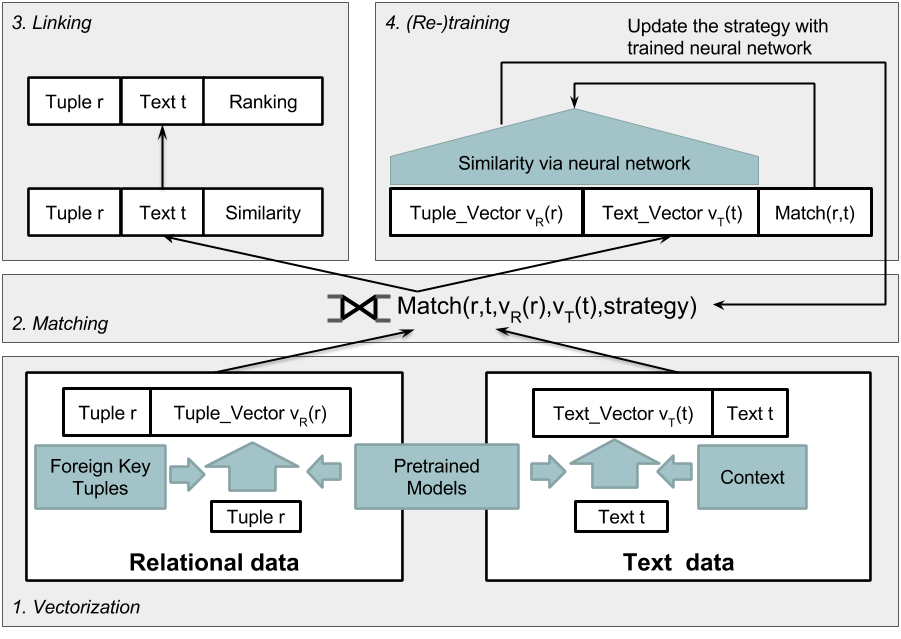}
	\caption{
		Architecture of IDEL (white rectangular boxes are data-containing
		components; colored figures represent system components).
		The workflow of the process consists of four steps:
		1) Vectorization: generate vector representations for both relational
		and text data;
		2) Matching: compute matching candidates from the relational data and
		the text by an initial query (bootstrapping) or by a previously
		learned neural network;
		3) Linking: rank and select the matching candidates with the highest
		similarities; and
		4) (Re-)training: the system uses matching candidates to (re-)train the neural
		network for improving models in next matching iterations.}
	\label{fig:architecture}
\end{figure*}

As a result, the following characteristics are realized in the IDEL architecture:
\begin{description}
  \item[Best of two worlds]
	  Users can seamlessly switch between SQL and Python, so that they can
		choose the best execution environment for each part of their data
		analytics.

  \item[Flexible and extensible]
	  IDEL provides a set of pre-trained neural network models.
	  In addition, it permits users to plug-in their own models or third-party
		models for entity linking.

  \item[Simple SQL-based user interface]
	  IDEL provides an SQL-\\based user interface to all parts of the system.
	  The whole workflow of entity linking can be executed by several calls to
		the implemented SQL UDFs.
	  All intermediate and final results can be stored in the underlying
		database for further analysis.

  \item[Robust to language errors]
	  IDEL adopts state-of-the-art neural embeddings for entity linking, which
		can achieve much higher precision under the four typical error sources
		(i.e. homonyms, hyponyms, synonyms and misspellings).
	  In addition, the system leverages extra information from the relational
		data, such as attribute values, and integrity constraints on data type
		and range.

  \item[No manual feature engineering]
	  IDEL does not require manual feature engineering; instead the system
		observes data distributions in the text database to represent best
		entities in relational and text data.
\end{description}

This paper is further structured as follows.
Section~\ref{sec:architecture} presents the architecture of IDEL.
Section~\ref{sec:model} describes the neural embedding models for representing
 text and relational data.
Section~\ref{sec:implementation} describes the implementation of IDEL.
Section~\ref{sec:eval} reports our preliminary evaluation results using the
 WebNLG data set with ten different entity types and thousands of manually
 labeled sentences.
Finally we discusses related work in Section~\ref{sec:related} and conclude
 with future outlook in Section~\ref{sec:conclusion}.

\section{IDEL Architecture}
\label{sec:architecture}

Figure \ref{fig:architecture} depicts the architecture of IDEL.
We assume that relational data is already stored in an RDBMS according to
 its schema.
In IDEL, we also store text data and neural embeddings in the same RDBMS.
Text data is simply stored as a collection of strings (e.g. a table with a
 single string column).
In this way, users can manage and query relational data together with text data
 and neural embeddings.
Our approach for entity linking addresses \emph{both mapping directions}, i.e.
 text to relational data and vice versa.
The process of entity linking can be divided into four major steps:

\paragraph{Step 1: Vectorization.}
First, we compute the respective vector representations (i.e. {\small{\sf
 Tuple\_Vector v$_R$(r)}} and {\small{\sf Text\_Vector v$_T$(t)}}) for the two
 data sets.
To create these vectors, we choose not to learn a neural network ourselves, but
 adopt a pre-trained model instead.
From the machine learning community, there already exist well-trained networks
 that are known to be particularly suitable for this kind of work, e.g
 SkipThought~\cite{kiros2015skip}.
Further, we can enrich both vector representations with additional
 discriminative features we can derive from their respective data sets.
For tuple vectors, we can use additional constraints in the relational data,
 such as foreign keys.
For text vectors, we can use context from surrounding sentences.
How to enrich these vectors is discussed in Sections~\ref{sec:rel_embedding}
 and~\ref{sec:text_embedding}, respectively.

\paragraph{Step 2: Finding matching candidates.}
The next step is to find matching candidates for entities in relational data with  mentions in 
 text data. Assume a user enters an SQL query such as the following to link relational and
 text data shown in Figure~\ref{fig:textjoinexample}:
\begin{verb}
SELECT e.*, o.*
FROM EntityMention e, Building b
WHERE LINK_CONTAINS(e.Mention, b.name, \$Strategy) = TRUE
\end{verb}
This query joins {\small{\sf EntityMention}} and tuples of {\small{\sf
 building}} and evaluates if a name of a building  is contained in the entity
 mentions. In addition to entity mentions and building names, the function
 {\small{\sf LINK\_CONTAINS}} takes a third parameter {\small{\sf \$Strategy}} so
 that different strategies can be passed. So far, we support an exact match and, most important for this work, a semantic match strategy. When computing the matches for the first time, there is generally very little
 knowledge about the data distribution.
Therefore, we suggest bootstrapping an initial candidate pool.
For example, one can generate exact matches with a join between words in entity
 mentions and words in relational tuples describing those entities.
This strategy is inspired by Snowball~\cite{DBLP:conf/dl/AgichteinG00}.
The initial matches can be used in later steps, such as linking and retraining.
Other sources for matchings are gold standards with manually labeled data.
However, this approach is highly costly and time consuming, because it requires
 expensive domain experts for the labeling and hundreds, or preferably even
 thousands of matchings.

\paragraph{Step 3: Linking.}
Now, we create linkings of matching entities.
We interpret entity linking as a ranking task and assume that an entity mention in the text is given and we try to find the
 $k$ most likely entities in the database.
On the other hand,  we can also assume that an entity in the database is given
 and we are interested in the $k$ most likely entity mentions in the text for
 this entity.
This step uses the matching candidates found in the previous step and generates
 a ranking.
If the matching function used in step 2 returns similarity values, this steps
 will leverage that information to compute a ranking and select the top $N$
 best matchings for further use.
In case the matching function, e.g. {\small{\sf LINK\_CONTAINS}}, does not
 produce similarity values (possibly due to the chosen strategy), all pairs of
 matching candidates are regarded to have equal similarity and hence will be
 included in the result of the linking.

\paragraph{Step 4: Retraining.}
An initial matching strategy, such as bootstrapping of candidates
 from exact matches with a join between words in sentences and words in tuples, is often unable to detect difficult natural language
 features such as homonyms, hyponyms, synonyms and misspellings. To improve results of the initial matching step, the system conducts a retraining step for improving neural models of the  semantic similarity function.
Thereby, the system updates previous models with retrained neural networks and
 recomputes the matching step.
IDEL permits repeating the training - updating - matching circle, because the database might get changed due to
 updates, insertions and deletions. If those changes alter the distributions of the data, the systems triggers the neural network to be retrained with the new data so as to match the new entities reliably while using existing matching models for bootstrapping training data and reducing manual labeling efforts. In the next section, we will describe in details how training is done.

%


\section{Embedding Models}
\label{sec:model}
\paragraph{Entity linking and deep learning}  Since very recently, entity linking techniques based on deep learning methods have started to gain more interests. The first reason is a significantly improved performance on most standard data sets reported by  TAC-KBP \cite{tac2016} \cite{tac2017}. Secondly, deep learning does not require costly feature engineering for each novel domain by human engineers. Rather, a system learns from domain specific raw data with high variance. Thirdly, deep learning based entity linking with character- and word-based embeddings often can further save language dependent costs for feature engineering. Finally, deep learning permits entity linking as a joined task of named entity recognition and entity linking \cite{DBLP:conf/coling/ArnoldDL16}, with complementary signals from images, tables or even from documents in other languages \cite{tac2017}. These recent findings triggered a move of the entire community to work on entity-linking with deep learning.

\begin{figure}
	\center
	\includegraphics[width=\columnwidth]{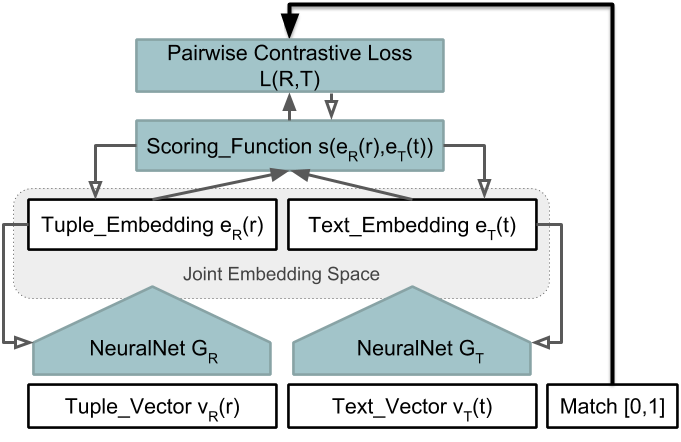}
	\caption{
		Overview of representing and matching entities in a joint embedding space
		in IDEL (white rectangular boxes are data-containing components;
		colored figures represent system components).
		First, we represent entities in relational tuples and in text
		data as vector representations. Second, the system generates a match
		between relational and text data. These candidates embeds the system
		with two feed forward network in the same vector
		space. We learn these networks into the same vector space by using a
		scoring- and a pairwise loss function. At prediction time,
		this scoring function measures the similarity between entities in the joint embedding space.}
	\label{fig:model}
\end{figure}

Figure~\ref{fig:model} gives an overview of our methods for representing and
 matching entities in a \emph{joint embedding space}.
It zoom in on the (re-)training step in the IDEL architecture (Figure~\ref{fig:architecture}).
In this section, we first provide a formal description for our transformation
 of relational and text data into their respective vector representations. Next, we formalize a joint embedding space, in which similar pairs of entities
 in the relational database and their corresponding entity mentions are kept
 close to each other, while dissimilar pairs further apart.
Then, we learn a common joint embedding space with a pairwise contrastive
 ranking loss function. Finally, in this joint space we compute a similarity between an embedding
 vector for relational and text data.

\subsection{Relational Data Embeddings}
\label{sec:rel_embedding}

\paragraph{Integrate various signals from the relational model in a single entity embedding.} The relational model features many rich signals for representing entities, such as relation and attribute names, attribute values, data types, and functional dependencies between values. Moreover, some relations may have further inter-dependencies via foreign keys. These relation characteristics are important signals for recognizing entities. Our approach is to represent the ``relational entity signatures'' relevant to the same entity in a single entity embedding.

\paragraph {Vector representation of entity relations.} To create embeddings we require a vector space representation. Therefore, we transform relations into vectors as follows:

Let $R\left(A_{1},\ldots,A_{n},FK_{1},\dots,FK_{m}\right)$ be a relation
with attributes $A_{1},\ldots,A_{n}$ and foreign keys $FK_{1},\dots,FK_{m}$
referring to relations $R_{FK_{1}},\dots,R_{FK_{m}}$. We define the domain of $R$ as $dom\left(R\right)=dom\left(A_{1}\right) \times \ldots \times dom\left(A_{n}\right) \times dom\left(FK_{1}\right) \times \ldots \times dom\left(FK_{m}\right)$.

\paragraph{Embed attribute data types.} Another important clue is the data type: we transform text data from alpha-numeric attribute values, such as {\small{\sf CHAR}}, {\small{\sf VARCHAR}} and {\small{\sf TEXT}}, in neural embeddings represented by the function $text2vec: TEXT \rightarrow \mathbb{R}^{m}$; we normalize numerical attribute values, such as {\small{\sf INTEGER}} and {\small{\sf FLOAT}}, with their mean and variance with the function $norm: \mathbb{R} \rightarrow \mathbb{R}$; and we represent the remaining attributes from other data types as a one-hot encoding (also known as 1-of-k Scheme) ~\cite{Bishop:2006:PRM:1162264}.
Formally, $\forall a_{i}\in A_{i}$ we define a vector $v(a)$ of $a$
as:

\[
v_{_{A_{i}}}(a_{i})=\begin{cases}
text2vec\left(a_{i}\right) & dom\left(A_{i}\right)\subseteq Text\\
norm\left(a_{i}\right) & dom\left(A_{i}\right)\subseteq Numbers\\
onehot\left(a_{i},A_{i}\right) & else
\end{cases}
\]

\paragraph{Embed foreign key relations.} Foreign key relations are another rich source of signals for representing an entity. Analogous to embeddings of entity relations from above, we encode embeddings for these relations as well. We define $\forall fk_{j}\in FK_{j}$ the vector $v_{_{FK_{j}}}\left(fk_{j}\right)$
of $fk_{j}$ as the sum of the vector representations of all foreign key tuples $v_{R_{FK_{j}}}\left(r_{fk_{j}}\right)$
\[
v_{FK_{j}}\left(fk_{i}\right)=\underset{r_{fk_{j}}\in R_{FK_{j}}}{\sum}v_{R_{FK_{j}}}\left(r_{fk_{j}}\right)
\]
 where $r_{fk_{j}}\in R_{FK_{j}}$is a foreign tuple from $R_{FK_{j}}$with
$fk_{j}$ as primary key.

\paragraph{Concatenating signature embeddings.} Finally, we concatenate all individual embeddings, i.e. entity relation, data type relation and foreign key relation, into a single embedding for each entity, i.e. $\forall r=\left(a_{1},\dots,a_{n},fk_{1},\dots,fk_{m}\right)\in R$, the vector $v_{R}(r)$ of tuple $r$ is defined as:
\[
v_{R}\left(r\right)=\begin{array}{c}
v_{A_{1}}\left(a_{1}\right)\oplus\dots\oplus v_{A_{n}}\left(a_{n}\right)\oplus\\
v_{FK_{1}}\left(fk_{1}\right)\oplus\dots\oplus v_{FK_{m}}\left(fk_{m}\right)
\end{array}
\]

\subsection{Text Embeddings}
\label{sec:text_embedding}

\paragraph{Representing entities in text as spans.}
Text databases, such as  INDREX \cite{kilias2015indrex} and System-T \cite{chiticariu2010systemt}, represent entities in text data as so-called \emph{span data type}:

\emph{Given a relation $T(Span,Text,Text)$ which contains tuples
$t=\left(span_{entity},text_{entity},text_{sentence}\right)$
where $span_{entity}$ $\in$ $Span$ is the span of the entity, $text_{entity}$ $\in$ $Text$
is the covered text of the entity and $text_{sentence}$ $\in$ $Text$ is
the covered text of the sentence containing the entity.}

The above formalization covers the entity name, the context in the same sentence and long range-dependencies in the entire in-document context of the entity. Thereby it implements the notion of \emph{distributional semantics} \cite{harris1954distributional}, a well-known concept in computational linguistics.

\paragraph{Vectorizing text spans and their context.}
Next, we need to vectorize spans and their context from above. We define the vectorization of text attributes of relations as a function $text2vec$ which can be ``anything'' from a pre-trained sentence embedding or a trainable recurrent network. In our model, we choose the popular and well suited approach \emph{SkipThought} \cite{kiros2015skip} from the machine learning community. Our rational is the following. First, SkipThought is based on unsupervised learning of a generic, distributed sentence encoder, hence there is no extra human effort necessary. Second, using the continuity of text in a document, SkipThought trains an encoder-decoder model that tries to reconstruct the surrounding sentences of an encoded passage. Finally, a SkipThought embedding introduces a semantic similarity for sentences. This can help with paraphrasing and synonyms, a core problem in resolving entities between relational and text data. In our implementation, we use the pre-trained sentence embeddings from SkipThought.

\subsection{Joint Embedding Space}

\label{sec:joint_embedding_space}
After the vectorization, we compute transformations for the entity-mentions and relational data embeddings in a joint embedding space in which pairs of entities in the relational database are already represented.  In this space similar entity mentions are placed close to each other while dissimilar pairs far apart.

Let the first transformation $e_{R}:R\rightarrow\mathbb{R}^{m}$ to compute an embedding for a tuple $r\in R$, while the second transformation  $e_{T}:T\rightarrow\mathbb{R}^{m}$ to compute an embedding for text $t\in R_{T}$. We define our transformations as follows:
\[
	\begin{array}{c}
		e_{R}\left(r\right)=G_{R}\left(v_{R}\left(r\right),W_{R}\right)\\
		e_{T}\left(t\right)=G_{T}\left(v_{T}\left(t\right),W_{T}\right)
	\end{array}
\]
where $ G_{R} $ denotes a neural network with weights $ W_{R} $ for the transformation of relational data and $ G_{T} $ a neural network with weights $ W_{T} $ for the transformation of text data. Weights $ W_{R} $ and $ W_{T} $ are learnable parameters and will be trained with Stochastic Gradient Decent. 

Depending on the vector representations used, $G_{R}$ and $G_{T}$ can be feed-forward, recurrent or convolutional neural networks, or any combination of them. In our implementation, we use feed-forward networks, because we transform the attribute values  of the relational data and the text with existing neural network models into a common vector representation.

\begin{figure*}
	\center
	\includegraphics[width=0.8\textwidth]{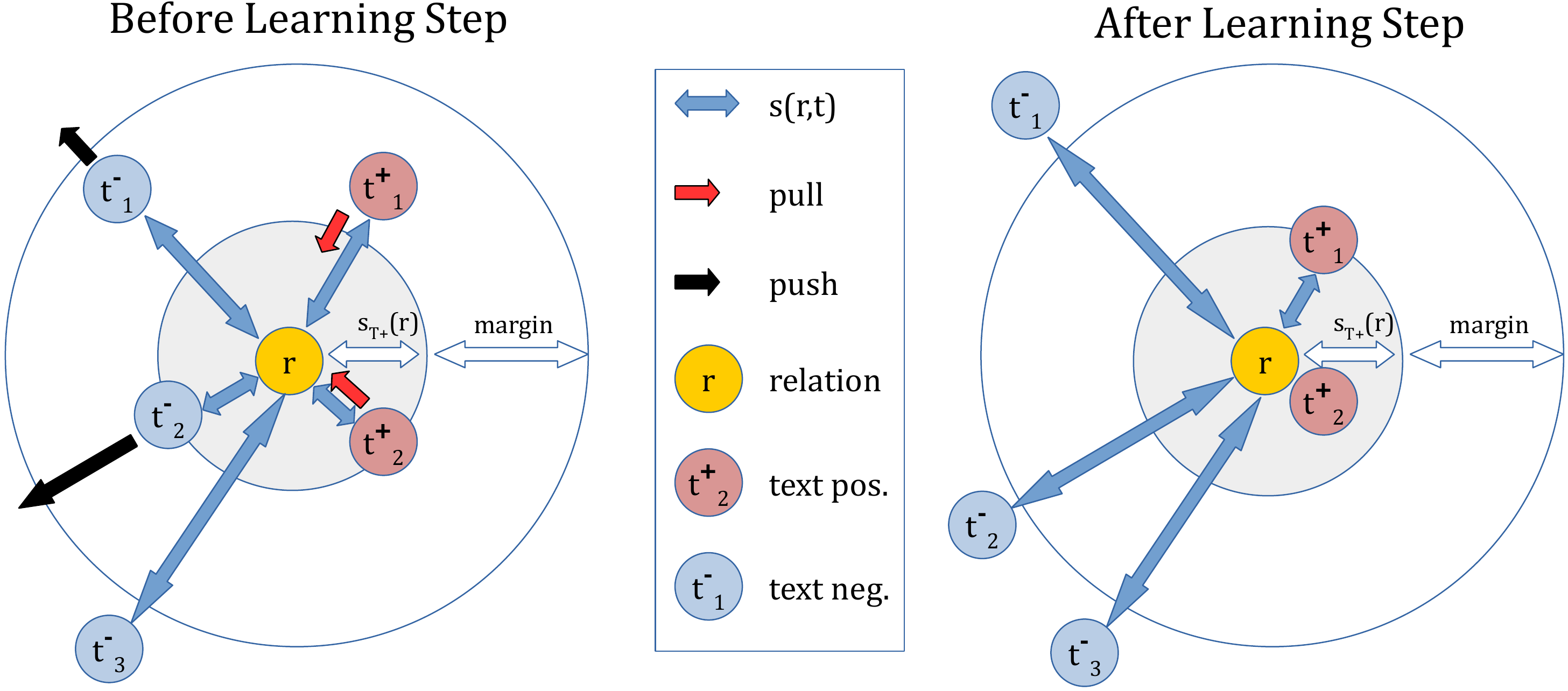}
	\caption{
		Overview of a learning step for a relation $r$ with a pairwise contrastive loss function. The left figure shows the state before the learning step, while the right figure shows the state after the training step. The relation $r$ is located in the center of each figure.  It is surrounded by matching text examples $t^+_1$, $t^+_2$ and not matching (i.e. contrastive) text examples $t^-_1$, $t^-_2$, $t^-_3$. The inner circle represents the average score between $r$ and  matching text $s_{T_{r}^{+}}\left(r\right)$ from equation \ref{eq:average_t}. The outer circle is the margin $m$. The loss function pulls matching text examples to the relation $r$ and pushes contrastive examples towards the outer circle.}
	\label{fig:loss}
\end{figure*}
\subsection{Pairwise Contrastive Loss Function}
\paragraph{Scoring function.}
By nature, text and relational embeddings represent different areas in a vector space created from our feed forward networks. Therefore, we must define a scoring function to determine how similar or dissimilar two representations in this vector space are. We compare these two embeddings with a scoring function $s\left(e_{R},e_{T}\right):\mathbb{R}^{m}\times\mathbb{R}^{m}\rightarrow\mathbb{R}_{\geq0}$,
where small values denote high similarities and larger values dissimilar entities.

Currently, we use the cosine distance as the scoring function $s\left(e_{R},e_{T}\right)$, since our experiments with  different distance measures, such as euclidean distance, show no notable effect on the accuracy of our results.

\paragraph{Loss function.} To train relational and text embeddings, $e_{R}$ and $e_{T}$, we use a Stochastic Gradient Decent, which conducts a backwards propagation of errors. Our loss (i.e. error) function is a variation of the \emph{pairwise contrastive loss}~\cite{kiros2014unifying} applied first to the problem of mapping images and their captions into the same vector space.
This loss function has desirable characteristics to solve our problem. 
First, it can be applied to classification problems with a very large set of classes, hence a very large space of 1000s of different entities. Further, it is able to predict classes for which no examples are available at training time, hence it will work well in scenarios with frequent updates without retraining the classifier. Finally, it is discriminative in the sense that it drives the system to make the right decision, but does not cause it to produce probability estimates which are difficult to understand during debugging the system. These properties make this loss function an ideal choice for our entity linking problem.

Applied to our entity linking task we consider either 1-N or M-N mappings between relations. We define our loss function as follows:

\begin{equation}
L\left(R,T\right)=\underset{r\in R}{\sum}L_{R}\left(r\right)+\underset{t\in T}{\sum}L_{T}\left(t\right)
\end{equation}
with $ L_{R}\left(r\right) $ as  partial loss for $ r $ and $ L_{T}\left(t\right) $ for $ t $
\begin{equation}
\label{eq:loos_r}
L_{R}(r)=\underset{t^{-}\in T_{r}^{-}}{\sum}
		max\{
			0,
			m
			+s_{T_{r}^{+}}(r)
			-s(e_{R}(r),e_{T}(t^{-}))\}
\end{equation}
\begin{equation}
\label{eq:loos_t}
L_{T}(t)=
	\underset{r^{-}\in R_{t}^{-}}{\sum}
		max\{
		0,
			m
			+s_{R_{t}^{+}}(t)
			-s(e_{T}(t),e_{R}(r^{-}))\}
\end{equation}
where $T_{r}^{-}$ denotes the set of \emph{contrastive} (i.e. not matching) examples and
$T_{r}^{+}$ denotes the set of matching examples of $T$ for $r$, such as respectively
$R_{t}^{-}$ and $R_{t}^{+}$ for $t$. The hyper-parameter margin $ m $ controls how far apart matching (positive) and not matching (negative) examples should be. Furthermore, the functions $ s_{R_{t}^{+}}\left(t\right) $ and $ s_{T_{r}^{+}}\left(r\right) $ calculate the average score of all positive examples for $ r $ and $ t $:
\begin{equation}
s_{R_{t}^{+}}\left(t\right)=
\frac{1}{\left|R_{t}^{+}\right|}\underset{r^{+}\in R_{t}^{+}}{\sum}s\left(e_{T}\left(t\right),e_{R}\left(r^{+}\right)\right)
\end{equation}
\begin{equation}
\label{eq:average_t}
s_{T_{r}^{+}}\left(r\right)=
\frac{1}{\left|T_{r}^{+}\right|}\underset{t^{+}\in T_{r}^{+}}{\sum}s\left(e_{R}\left(r\right),e_{T}\left(t^{+}\right)\right)
\end{equation}

Figure \ref{fig:loss} shows a learning step of this loss function for a relation $r$. In equations \ref{eq:loos_r} and \ref{eq:loos_t}, the addition of $s_{R_{t}^{+}}\left(t\right)$ and $s_{T_{r}^{+}}\left(r\right)$ pulls embeddings vectors for positive examples together during the minimization of the loss function by decreasing their score. Contrarily, the subtraction for a contrastive example of $s\left(e_{T}\left(t\right),e_{R}\left(r^{-}\right)\right)$ and $s\left(e_{R}\left(r\right),e_{T}\left(t^{-}\right)\right)$ pushes embedding vectors further apart, because increasing their score minimizes this subtraction. The margin limits the score for a contrastive example, since there the loss function cannot push the embedding vector of a contrastive example further. This is crucial to learn mappings between two different vector spaces.

Overall, we are not aware of any other work where loss functions for mapping pixels in images to characters are  applied to the problem of linking entities from text to a table. IDEL is the first approach that abstracts this problem to entity linking. For our specific problem we therefore modified  the loss function of \cite{kiros2014unifying} by replacing a single positive example with the average score of all positive examples $s_{R_{t}^{+}}\left(t\right)$ or $s_{T_{r}^{+}}\left(r\right)$. This pulls all positive examples together, enabling our loss function to learn 1-N and M-N mappings between relational data and text.


\subsection {Hyper-parameters and Sampling}
\paragraph{Hyper-parameters.} We evaluate several configurations for representing neural networks with relational $G_{R}$ or text data $G_{T}$. Our representation for relational data contains three layers: the input layer containing 1023 neurons, the second layer 512 neurons and the output layer 256 neurons. For representing text embeddings $G_{T}$ we use two layers: an input layer with 1024 neurons and an output layer with 256 neurons. We choose fewer layer for $ G_{T} $, because the dimensionality of their input is smaller than that of the relational data. All layers use the fast and aggressive activation function \emph{elu}. We train the model via gradient descent with Adam as optimizer and apply dropout layers to all layers with a keep probability of $0.75$. Since we use dropouts, we can choose a higher learning rate of $1\mathrm{e}{-05}$ with an exponential decay of $0.9$ every $1000$ batches and set our margin to 0.001.

\paragraph{Sampling batches from training set}
\label{par:sampling}
Our training set consists of two tables (relational and text data) with a binary mapping between their tuples. A tuple here denotes an entity represented as span from text data, including span and context information, and an entity, including attribute values, from a relational table. We train our model with batches. Entities in training data, both text and relational, are often Zipfian-distributed, hence, very few popular entities appear much more frequently than the torso and long tail of other entities. Analogous to Word2Vec in \cite{mikolov2013distributed}, we compensate this unbalanced distribution during sampling batches for our training process. Our sampling strategy learns a distribution of how often our classifier has seen a particular matching example for an entity in prior batches. Based on this distribution, we draw less frequently seen examples in the next batches and omit frequently seen entity matching examples. Moreover, true positive and true negative matchings are unbalanced. Because of the nature of the entity linkage problem, we see much more negative examples than positive examples. To compensate this additional imbalance, each training example contains at least one true positive matching example (an entity represented in both text and relational data) and other negative example.

\section{Implementation}
\label{sec:implementation}

\begin{figure}
	\centering
	\includegraphics[width=\columnwidth,clip,trim=1.5cm 3.5cm 1.5cm 5.5cm]{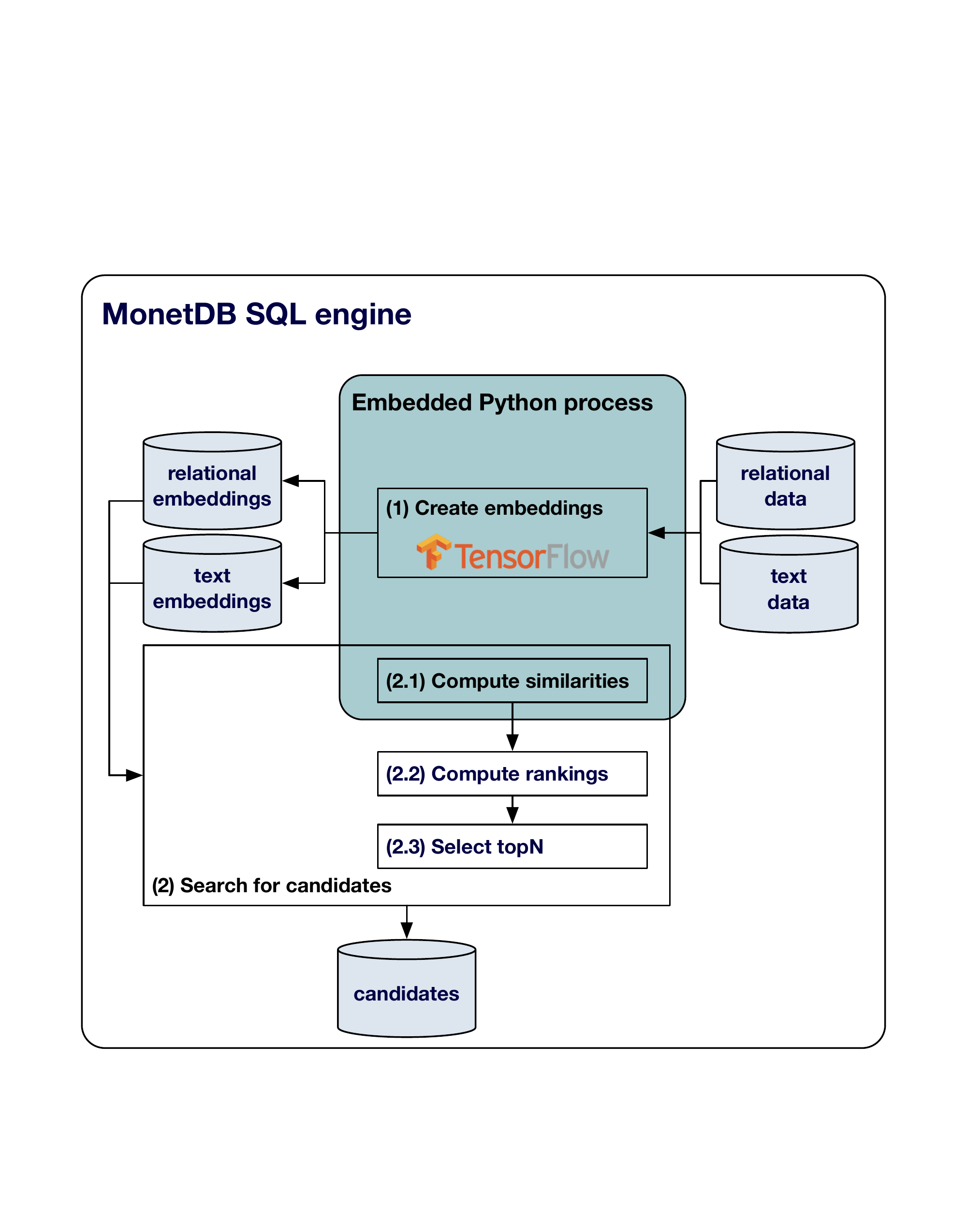}
	\caption{
		System architecture of IDEL, which supports an entity linking process
		 by first creating embeddings of relational and text data, and then
		 searching for candidates, which is currently done by computing first
		 the similarities, then the rankings, before selecting the top N
		 candidate. 'Search for candidates' employs the Nearest-Neighbor-Search index Spotify Annoy
        for fast retrieval of top-n matching entities.
	}
	\label{fig:processandarch}
\end{figure}
Despite the fact that several technologies are investigated by the computational linguistics or the machine learning community, no other database system currently permits the execution of neural text mining in the database. One reason is the choice of the RDBMS, which has a significant impact on the overall system. The ability of MonetDB to support in-database-analytics through user defined Python/SQL functions makes it possible to integrate different types of
systems to solve the overall entity-linkage task. We briefly describe this integration in this section. 

\subsection{MonetDB/Python/TensorFlow/Annoy}
We have integrated the entity linking process into a single RDBMS, MonetDB, as depicted in Figure~\ref{fig:processandarch}, and store text, relational data and embeddings in MonetDB. The computation is either implemented in SQL queries or SQL UDFs in Python. MonetDB is an open-source columnar RDBMS optimized for in-memory  processing of analytical workloads~\cite{DBLP:journals/cacm/BonczKM08}.
In recent years, MonetDB enriched its support for in-database analytics by,
 among others, introducing MonetDB/Python integration through SQL
 UDFs~\cite{Raasveldt:2016:VUC:2949689.2949703}. As a result, MonetDB users can specify Python as the implementation language
 for their SQL UDFs. Basically, any Python libraries accessible by the MonetDB
 server can be imported. In Our work we base on the deep learning library TensorFlow\footnote{https://github.com/tensorflow} and the nearest neighbor search index for neural embeddings, Spotify Annoy \footnote{https://github.com/spotify/annoy}.
When such an \emph{SQL Python UDF} is called in an SQL query, MonetDB 
 automatically starts a Python subprocess 
 to execute the called UDF. MonetDB exchanges data of relational tables between the SQL engine and the
 embedded Python process by means of NumPy arrays.

The MonetDB/Python integration features several important optimizations to allow efficient executions of SQL Python UDFs:

\begin{itemize}
	\item \emph{Zero data conversion cost.}
		Internally, MonetDB stores data of each column as a C-array.
		This binary structure is the same as the NumPy arrays, hence, no data
		 conversion is needed between SQL and Python.
	\item \emph{Zero data transfer cost.}
		The MonetDB SQL engine and the embedded Python process share the same
		 address space. 
		Passing columnar data between the two systems merely means passing
		 pointers back and forth.
	\item \emph{Parallel execution of SQL Python UDFs.}
		When processing an SQL query, MonetDB can speed up the execution by
		 splitting up columns and executing the query on each part in
		 parallel.
		MonetDB/Python integration allows an SQL Python UDF to be declared as
		 parallelizable, so that this UDF can be part of a parallel query
		 execution plan (otherwise the UDF will be treated as a blocking
		 operator).
\end{itemize}

Figure~\ref{fig:processandarch} shows the implementation of IDEL in
 MonetDB.
We store relational data  in MonetDB according to their schemas and  text data in a table with a single string-typed column.
First, we create embedding vectors for both relational and text  data by two SQL Python UDFs, one for each input table. This step leverages TensorFlow's machine learning features to load the  pre-trained neural network and apply it on the input tables. We return embedding vectors as NumPy arrays and store them as BLOBs. The second step finds matching candidates with the highest similarities among embeddings. We employ nearest neighbor search with Annoy for a given embedding, compute a ranked list for each entity according to their similarities and finally return TopN candidates. All steps are implemented in SQL. 


 
Changes inside this architecture, such as different embedding models or similarity functions, are transparent to upper layer applications.

\subsection{Create Embeddings}
\label{sec:pythonudfs}


We abstract core functionalities for entity linking as SQL UDFs. This design principle permits us exchanging functionalities  in UDFs for different data sets or domains.


\paragraph{UDF:EmbedSentences}
This UDF embeds text sentences into the joint vector space.
It applies $v_{T}$ and $G_{T}$ from a trained model to generate $e_{T}$. Because of the learned joint vector space, the function $G_{T}$ is coupled on $G_{T}$ which computes the tuple embedding for each table to which we want to link.
It takes as input a NumPy array of strings, loads a trained neural network model into main memory and apply this model to the NumPy array in parallel.
Due to the implementation of our model in TensorFlow, this UDF can leverage GPUs to compute the embedding vectors. Finally, this function transforms an array of embedding vectors  into an array of BLOBs and returns it to MonetDB. The following example executes the UDF \emph{$embed\_sentence\_for\_building$} to retrieve sentences about buildings and return their embeddings.
\begin{verb}
CREATE TABLE embedd_sentences_for_building AS
  SELECT *, embed_sentence_for_building(sentence) as embedding
  FROM sentences;
\end{verb}
This UDF returns a BLOB in Python and executes a Python script that encodes each sentence into an embedding. 
\begin{verb}
CREATE FUNCTION embed_sentences_building(sentences STRING) 
RETURNS BLOB LANGUAGE PYTHON
{
	from monetdb_wrapper.embed_udf import embed_udf
	return embed_udf().run("path/to/repo","path/to/model",
				"sentences", {"sentences": sentences })
};
\end{verb}

 %

\paragraph{UDF:EmbedTuples}
This UDF  embeds tuples of a table into the joint vector space. It applies a trained neural network model on $v_{R}$ and $G_{R}$ to generate
 $e_{R}$. As input, it assumes arrays of relational columns, and loads and applies a trained model in parallel to input relations and outputs  embedding vectors as an array of BLOBs. The exact signature of this UDF depends on the
 schema of the table. In the following example, we encode the table \emph{building} with attributes {\small{\sf name}}, {\small{\sf address}} and {\small{\sf owner}} in the embedding. 
\begin{verb} 
CREATE TABLE building_with_embedding AS
  SELECT *, embed_building(name, address, owner) as embedding
  FROM building;
\end{verb}

\subsection{Search for Candidates}
\paragraph{UDF:QueryNN}  The next task is, given a vector (e.g. an entity represented in text), to retrieve a set of similar vectors (e.g. tuples representing this entity in a table). A naive search solution would execute a full table scan and compute for each pair the distance between the two vectors. With a growing numbers of entities in tables or text, this operation becomes expensive. Therefore, we represent embeddings for entities in a nearest neighbor search index for neural embeddings. 
Following  benchmarks of \cite{DBLP:conf/sisap/AumullerBF17}, we implemented this index with Spotify Annoy  for four reasons. First, Annoy is almost as fast as the fastest libraries in the benchmarks. Second,  it has the ability to use static files as indexes which we can share across processes. Third, Annoy decouples creating indexes from loading them and we can  create indexes as files and map them into memory quickly. Finally, Annoy has a Python wrapper and a fast \Cpp kernel, and thus fits nicely into our implementation.

The index bases on random projections to build up a tree. At every intermediate node in the tree, a random hyperplane is chosen, which divides the space into two subspaces. This hyperplane is chosen by sampling two points from the subset and taking the hyperplane equidistant from them. Annoy applies this technique \emph{t} times to create a forest of trees. Hereby, the parameter \emph{t} balances between precision and performance, see also work on Local sensitive hashing (LSH) by \cite{DBLP:conf/stoc/Charikar02}. During search, Spotify Annoy traverses the trees and collects \emph{k} candidates per tree. Afterwards, all candidate lists are merged and the TopN are selected. We follow experiments of \cite{DBLP:conf/sisap/AumullerBF17} for news data sets and choose $t=200$ for $k=400000$ neighbors for $N=10$.

The following examples executes a nearest neighbor search for an embedding representing  relational tuples of table building in the space of indexed sentences that represent an entity if type building. The query returns the Top10 matching sentences for this relational entity. 
\begin{verb}
SELECT *
FROM query_index((
       SELECT id, embedding, 10, 400000,
	          index_embedd_sentence_for_building
       FROM embedd_building)) knn,
    building_with_embedding r,
    sentences_for_building_with_embedding s
WHERE r.id = knn.query_key AND s.id = knn.result_key;
\end{verb}

\section{Experimental Evaluation}

\label{sec:eval}

\begin{table*}
               \centering
               \begin{tabular}{|l|l|l|l|l|l|l|l|}
                              \hline
                              Entity Family & Entity type     & Instances & Tuples & Sentences & Sentences/Instance & Columns & Avg. tuple density \\
\hline
                              Location      & Airport         & 67        & 662    & 2831      & 42.25      & 52      & 0.01       \\
\hline
                              Location      & Building        & 58        & 380    & 2377      &  40        & 46      & 0.10        \\
\hline
                              Location      & City            & 65        & 243    & 609       &  9.37      & 25      & 0.16               \\
\hline
                              Location      & Monument        & 10        & 111    & 783       &  78.3      & 31      & 0.17       \\
\hline
                              Product       & ComicsCharacter & 40        & 116    & 749      &  18.27     & 20      & 0.21       \\
\hline
                              Product       & Food            & 59        & 641    & 3646      &  61.8      & 34      & 0.14       \\
\hline
                              Product       & WrittenWork     & 52        & 486    & 2466      &  47.42     & 49      & 0.10         \\
\hline
                              Brand  & University      & 17        & 308    & 1112             &  65.41     & 39      & 0.16
\\ \hline
                              Brand  & SportTeam       & 55        & 471    & 1998             &  36.32     & 34      & 0.14
\\ \hline
                              Person        & Astronaut       & 17        & 459    & 1530      &  90        & 38      & 0.18       \\
\hline
\end{tabular}
               \caption{The WebNLG data set provides a manually labeled ground truth for 10 different entity types, such as products, locations, persons and brands. It features several thousands of manually annotated sentences. Moreover, its structured representation describes entities with at least 20 up to 52  attributes. However, most attribute values are populated only sparsely: in this table, ``Avg. tuple density'' denotes the portion of non-null attribute values for all entities of a particular type. Hence, most structured entity tuples have only a few attribute values different from NULL.}
               \label{tab:datasets}
\end{table*}


\subsection{Data Set Description}

\paragraph{WEBNLG} Authors  in~\cite{DBLP:journals/corr/Perez-Beltrachini17} provide an overview of data sets where entities have multiple attributes and are matched against text data (i.e. sentences). One of the largest manually labeled data sets is WebNLG~\cite{DBLP:conf/coling/Perez-Beltrachini16} with ten different entity types and thousands of manually labeled mappings from entity data in RDF to  text. To use WebNLG for our experiments, we have transformed the RDF representation  into a relational model and evaluated our work against it. WebNLG contains relevant entity types for our scenarios from the introduction, such as \emph{building}, comics character, food or written work (which are products) or universities and sport teams (which are brands).  Because of these different domains, this ``mix'' is particularly hard to detect  in single entity linking system and it realistically models data for our  example use case.

The following two examples from WEBNLG show candidate sentences for entities of the type `building'. For the first entity three attributes are shown in structured and text data, while the second example features five attributes. Note that the later attribute is described over multiple sentences. Attribute names are highly ambiguous and do not match words in texts. Furthermore, the position of attributes in text varies. 
%
\begin{verb}
<entry size="3" eid="Id24" category="Building">
  <modifiedtripleset>
    <mtriple>200_Public_Square | floorCount | 45</mtriple>
    <mtriple>
      200_Public_Square | location | "Cleveland, Ohio 44114"
    </mtriple>
    <mtriple>200_Public_Square | completionDate | 1985</mtriple>
  </modifiedtripleset>
  <lex lid="Id3" comment="good">
    200 Public Square, completed in 1985, has 45 floors and is
    located in Cleveland, Ohio 44114.
  </lex>
</entry>

<entry size="5" eid="Id1" category="Building">
  <modifiedtripleset>
    <mtriple>103_Colmore_Row | floorCount | 23</mtriple>
    <mtriple>103_Colmore_Row | completionDate | 1976</mtriple>
    <mtriple>103_Colmore_Row | architect | John_Madin</mtriple>
    <mtriple>
      103_Colmore_Row | location |
      "Colmore Row, Birmingham, England"
    </mtriple>
    <mtriple>John_Madin | birthPlace | Birmingham</mtriple>
  </modifiedtripleset>
  <lex lid="Id1" comment="good">
    103 Colmore Row is located on Colmore Row, Birmingham,
    England. It was designed by the architect, John Madin,
    who was born in Birmingham. It has 23 floors and was
    completed in 1976.
  </lex>
</entry>
\end{verb}

%
\paragraph{Quality of training data} Table~\ref{tab:datasets} gives an overview of the WebNLG data set and some
 important statistics for each \emph{Entity type}. \emph{Instances} denotes the number of distinct entities for each type in the
 relational data.
\emph{Tuples} denotes the number of tuples in the relational data for each
 distinct instance.
\emph{Sentences} denotes the number of sentences that contain at least one of
 these instances in text data.
\emph{Sentences/Instance} counts the average ratio of how often an instance is
 represented in a sentence.
In particular, for types {\small{\sf City}} and {\small{\sf ComicsCharacter}} we observe relatively very
 few sentences representing each entity.
As a result, the system might learn less variances during training and sampling
 for these data types.
\emph{Columns} denotes the number of distinct attributes in the relational
 schema for this entity type.
\emph{Avg. tuple density} denotes the proportion of an attribute with
 non-NULL values averaged over all tuples for this type.
We observe that all entity types feature sparsely populated
 attribute values only.
Some entity types are described with a rather large number of attribute values
 (up to 52), while most entity types contain $20 \sim 30$ attributes.
For example, the entity types {\small{\sf WrittenWork}} and {\small{\sf Airport}} are described
 with roughly 50 sparsely populated attribute values.

\paragraph{Training, test and cold start scenario} In realistic situations,  new entities are regularly added to the database. Hence, it is important for our system to recognize such \emph{cold start entity representations} without needing to be re-trained. Hence, we need to consider previously seen entities for which we learn new matchings (hot and running system) and entities we have never seen during training (cold start scenario). To simulate these two scenarios we choose the same setup as described in \cite{DBLP:conf/emnlp/GuptaSR17}  and split  the set of relational entity instances into 20\% \emph{unseen} entities for the cold start scenario. We  kept the remaining 80\% as previously seen entities and  split this set again into 80\% for training and 20\% for testing.

\begin{table*}
	\centering
	\begin{tabular}{|l|l|l|l|l|l|l|l|l|l|}
		\hline
		& \multicolumn{3}{l|}{Prec@1}    & \multicolumn{3}{l|}{Prec@5}    & \multicolumn{3}{l|}{Prec@10}   \\ \hline
		& test     & train    & unseen   & test     & train    & unseen   & test     & train    & unseen   \\ \hline
		Airport         & 0.90 & 0.98 & 0.54 & 0.96 & 0.99 & 0.70 & 0.99 & 1 & 0.79 \\ \hline
		Astronaut       & 0.91 & 0.96 & 0.88 & 0.97 & 0.99 & 0.98 & 0.98 & 1 & 0.98 \\ \hline
		Building        & 0.89 & 0.99 & 0.77 & 0.94 & 1    & 0.90 & 0.97 & 1 & 0.95 \\ \hline
		City            & 0.73 & 0.98 & 0.93 & 0.93 & 1    & 0.98 & 0.96 & 1 & 1    \\ \hline
		ComicsCharacter & 0.76 & 0.99 & 0.29 & 0.97 & 1    & 0.80 & 0.98 & 1 & 0.97 \\ \hline
		Food            & 0.85 & 0.94 & 0.69 & 0.94 & 0.98 & 0.90 & 0.94 & 0.98 & 0.91 \\ \hline
		Monument        & 0.94 & 1    & 0.90 & 0.98 & 1    & 0.98 & 1    & 1    & 1        \\ \hline
		SportTeam       & 0.90 & 1    & 0.66 & 0.97 & 1    & 0.83 & 0.99 & 1    & 0.92 \\ \hline
		University      & 0.95 & 1    & 0.93 & 0.99 & 1    & 1    & 1    & 1    & 1        \\ \hline
		WrittenWork     & 0.88 & 0.95 & 0.63 & 0.97 & 0.99 & 0.79 & 0.99 & 0.99 & 0.86  \\ \hline
	\end{tabular}
	\caption{Accuracy in Precision@k of the trained model for each entity type. In \emph{test} we observe for all entity types, except {\small{\sf City}} and {\small{\sf ComicsCharacter}}, a very high Precision@1\textgreater0.80. Even for a cold start scenario (columns \emph{unseen}) we can still observe decent Precision@1\textgreater0.60 for all entity types, except {\small{\sf Airports}} and {\small{\sf ComicsCharacter}}. In addition, we included columns Prec@5 and Prec@10 to illustrate that our system de facto can retrieve the correct matching entities at lower ranks. Recall is not shown in this diagram, since we measure precision numbers for all entities in our test data set.}
	\label{tab:accuracy}
\end{table*}

\begin{figure*}
	\centering
	\begin{subfigure}[t]{0.33\textwidth}
		\centering
		\includegraphics[width=\textwidth]{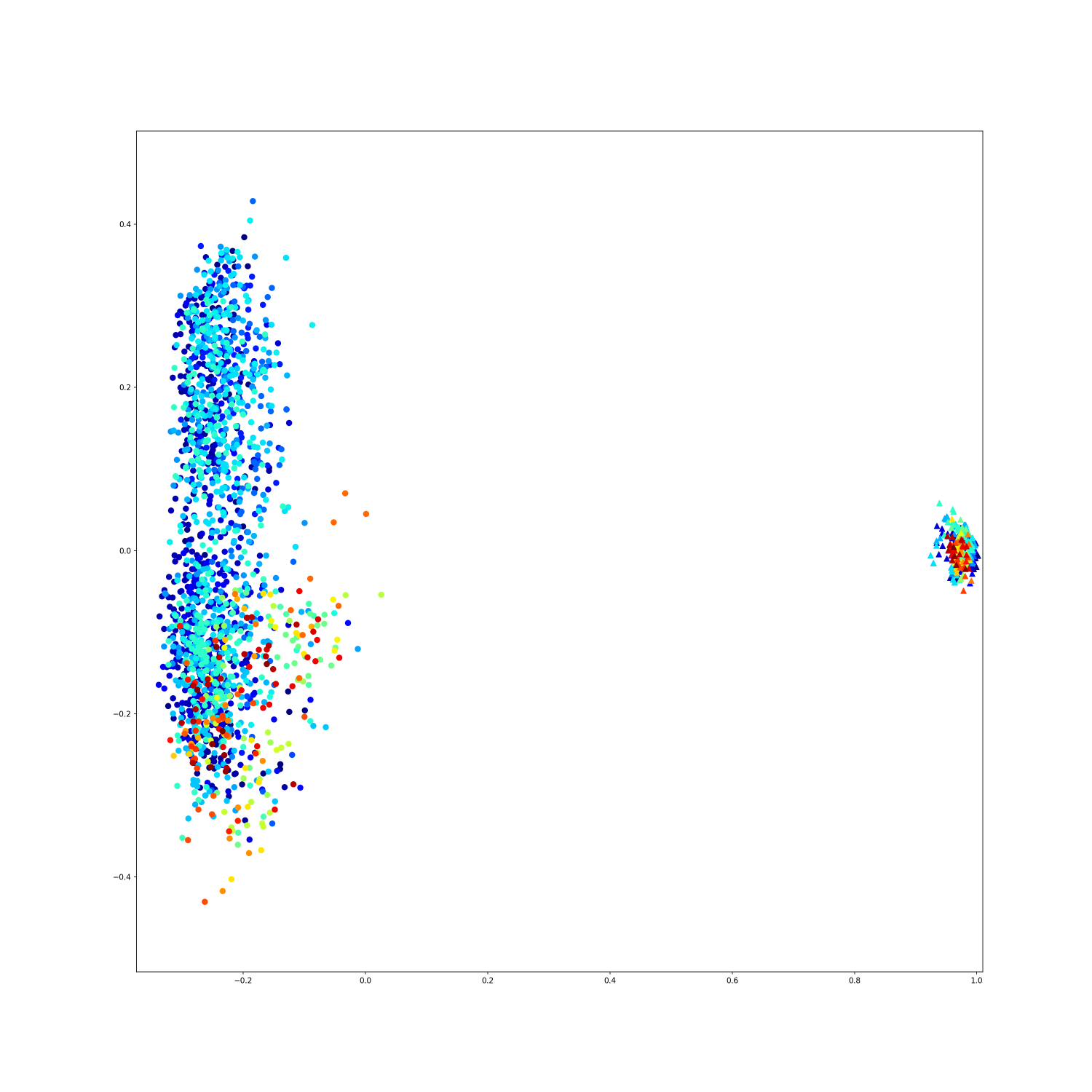}
		\caption{Initial}
	\end{subfigure}%
	\begin{subfigure}[t]{0.33\textwidth}
		\centering
		\includegraphics[width=\textwidth]{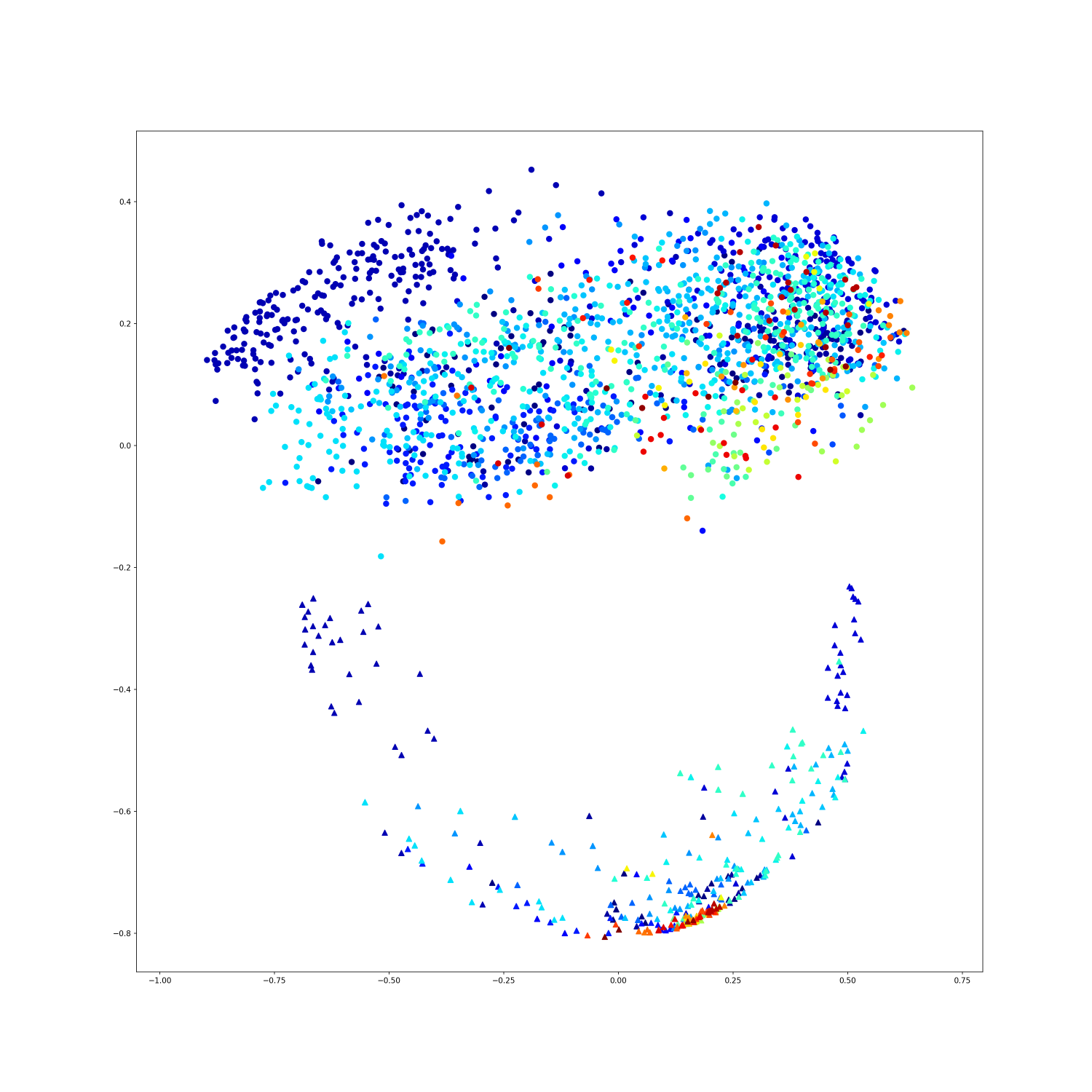}
		\caption{100 Batches}
	\end{subfigure}
	\begin{subfigure}[t]{0.33\textwidth}
		\centering
		\includegraphics[width=\textwidth]{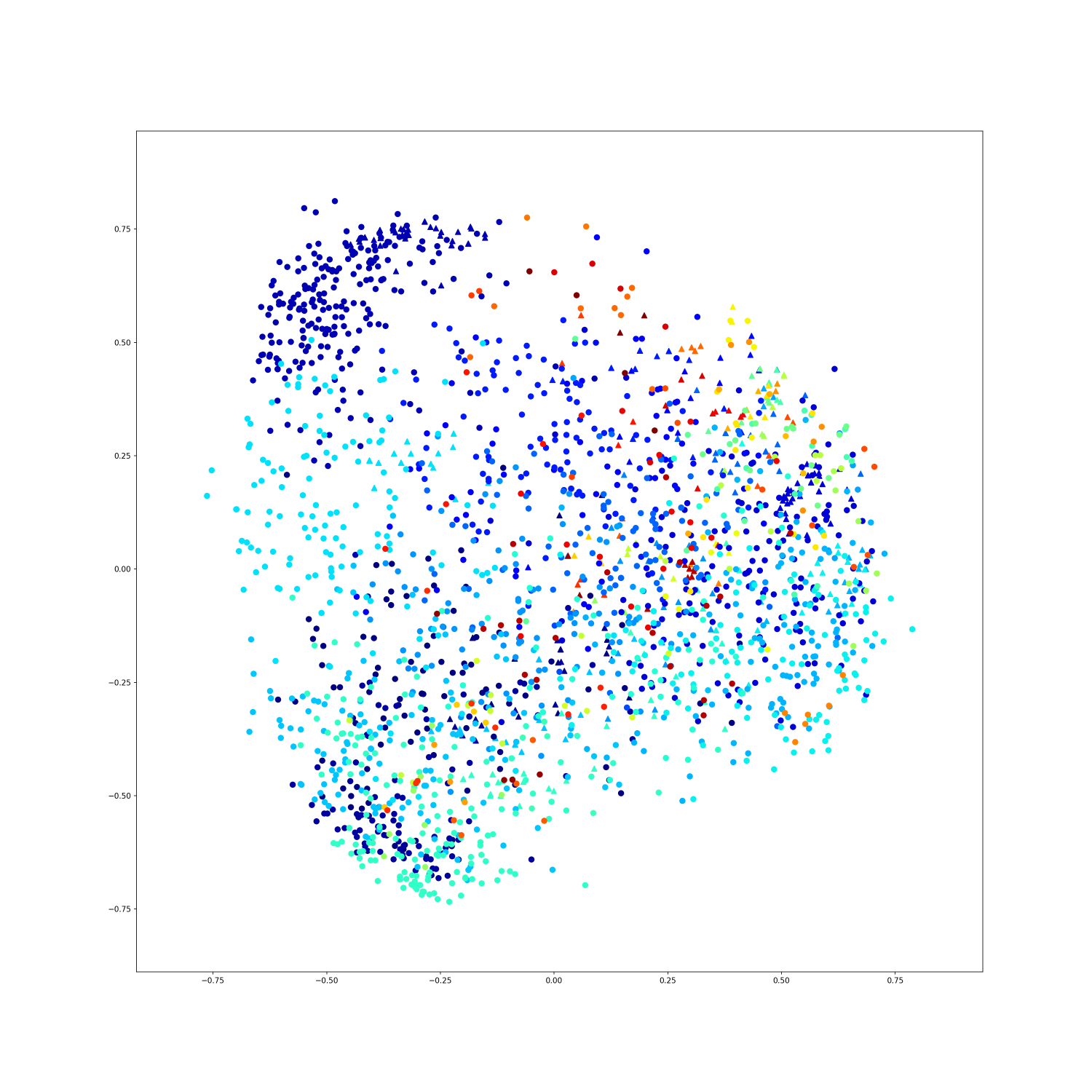}
		\caption{200 Batches}
	\end{subfigure}
	\\
	\begin{subfigure}[t]{0.33\textwidth}
		\centering
		\includegraphics[width=\textwidth]{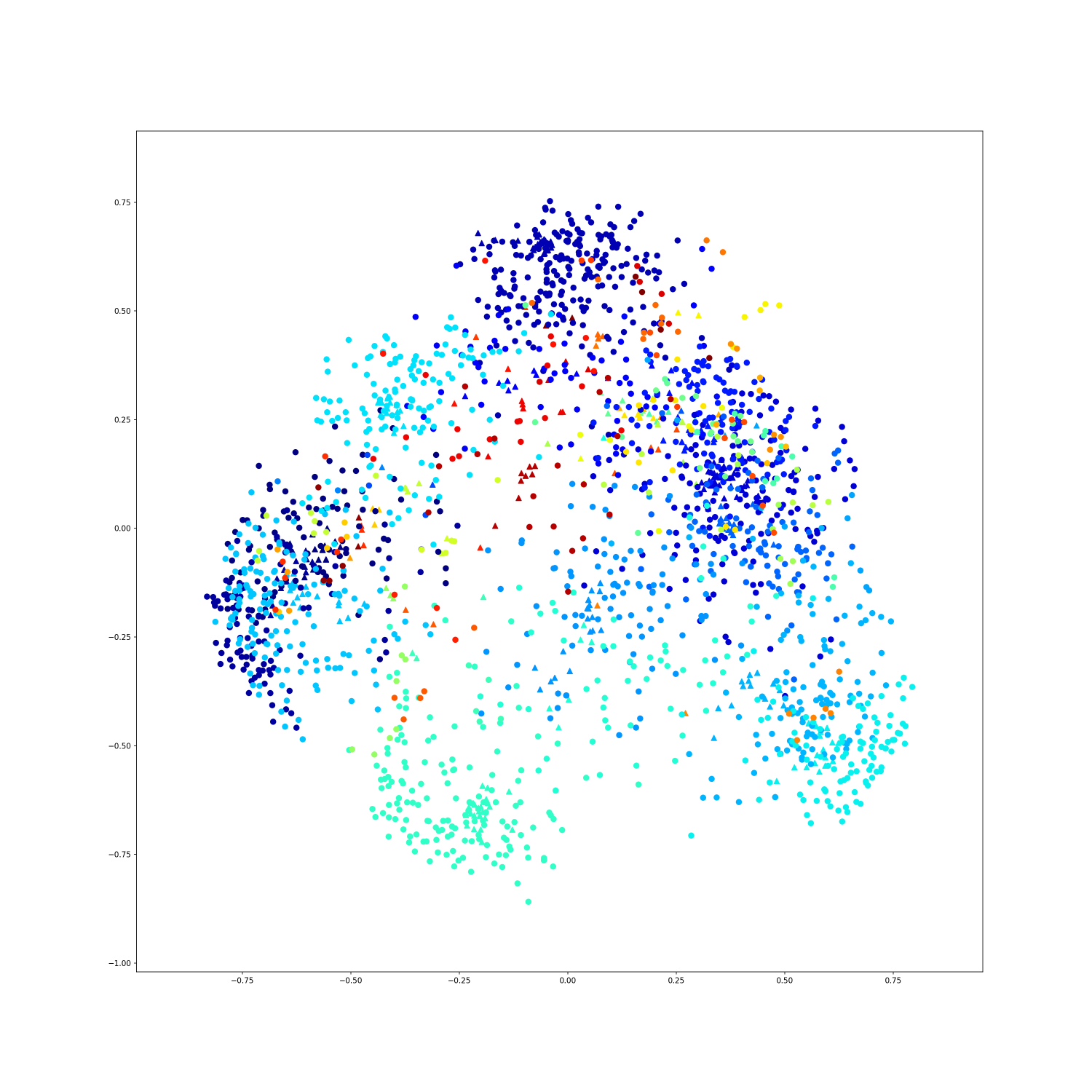}
		\caption{600 Batches}
	\end{subfigure}
	\begin{subfigure}[t]{0.33\textwidth}
		\centering
		\includegraphics[width=\textwidth]{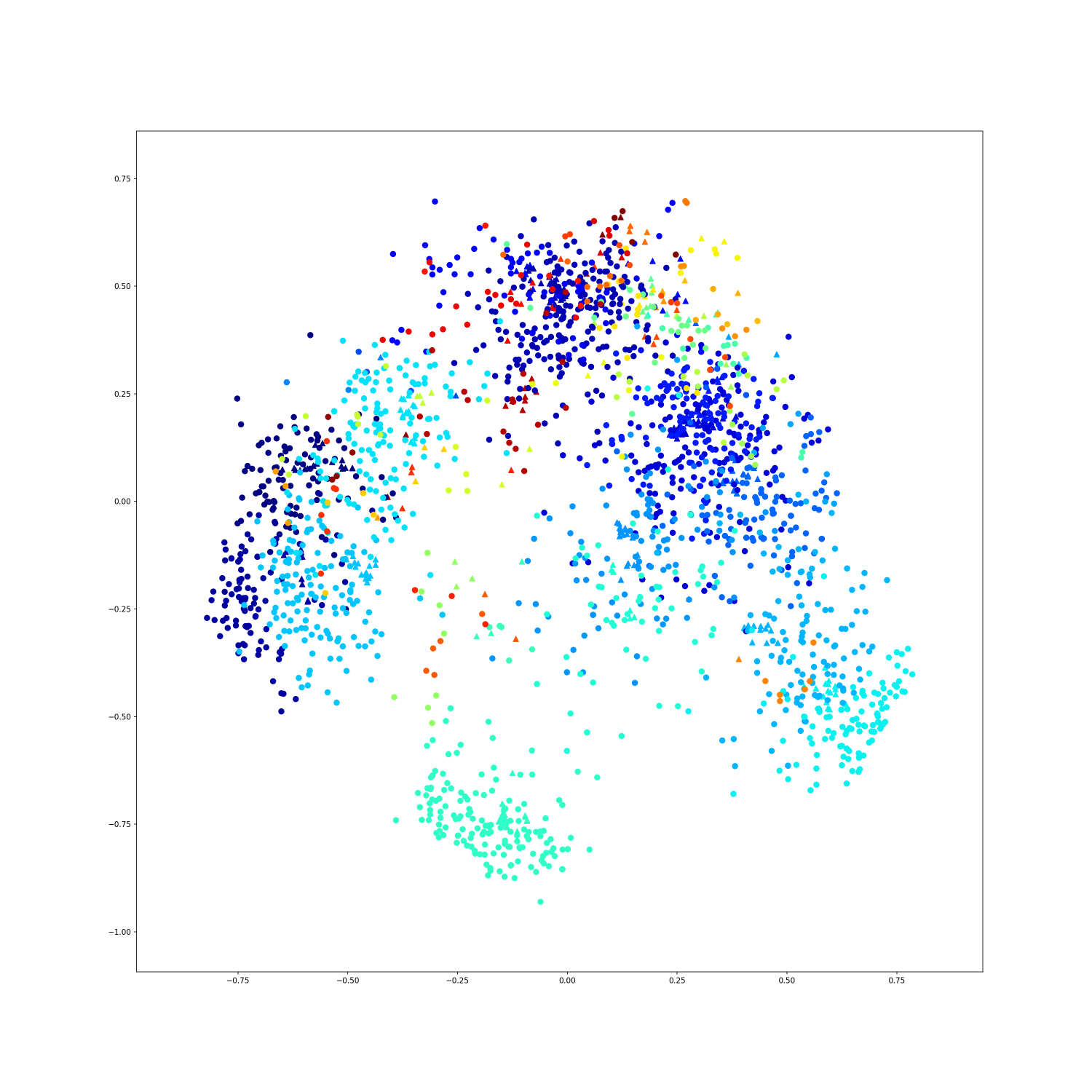}
		\caption{1200 Batches}
	\end{subfigure}
	\begin{subfigure}[t]{0.33\textwidth}
		\centering
		\includegraphics[width=\textwidth]{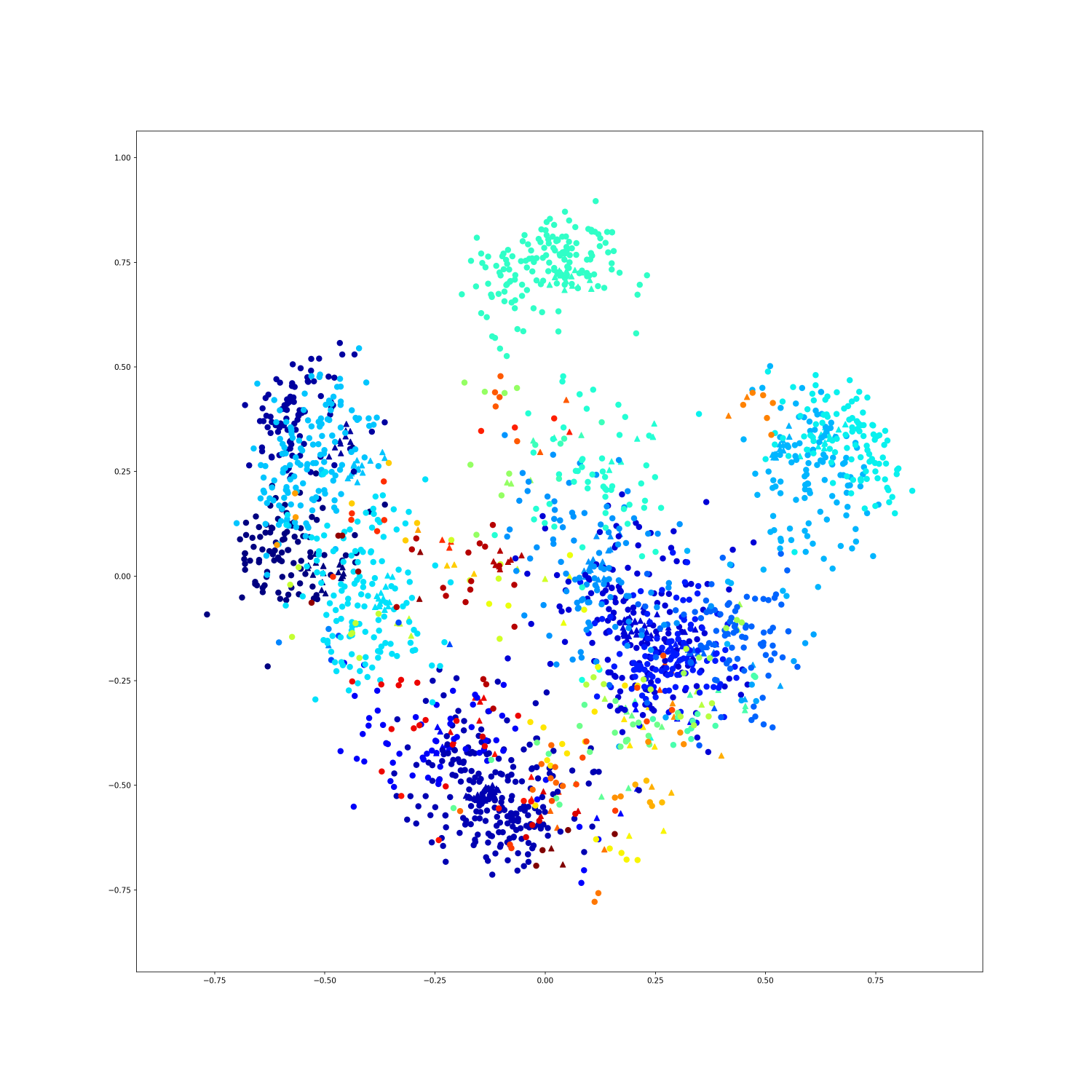}
		\caption{2400 Batches}
	\end{subfigure}
	\caption{The above six figures show the vector space of text and relational embeddings during the training period at varying batch samples. In figure (a) we observe two tuple clusters: one for text data and one for relational data. Different colors indicate different entities (here limited to 58 different entities of the class ``Building''). During each batch sample our method clusters similar entities in the latent space and separates dissimilar entities. Finally, in figure (f) we observe a system state with rather sharply clustered text and relational representation for light medium, dark blue and green entities. At this stage, red and yellow entities still need to be clustered and separated from others.}
	
	\label{fig:batches}
\end{figure*}

\subsection{Experimental Setup}
\paragraph{System setup}
We implemented our model using TensorFlow 1.3, NumPy 1.13 and integrated it into MonetDB (release Jul2017-SP1). We installed these software packages on a machine with two Intel\textsuperscript{\textregistered} Xeon\textsuperscript{\textregistered} CPUs E5-2630 v3 with 2.40GHz, 64 GB RAM, and SSD discs.



\paragraph{Measurements}
Our first set of experiments measures the effectiveness of our embeddings and entity linking. Given an entity representation from the relational data, the output of entity linking is an ordered list of sentences where this entity likely appears. A common measure is \emph{Precision@1}, which counts how often the system returns the correct sentence at rank 1. Analogously, \emph{Precision@5} and \emph{Precision@10} count how often the correct result returned by the system is among the first five and ten results.
Our second set of experiments measures the efficiency. We measure execution times for loading and creating embeddings before query run time and for generating candidates, executing the similarity measure on candidates and ranking candidates at runtime.
\subsection{Experimental Results}
\paragraph{Entity Linking with very high precision}
Table~\ref{tab:accuracy} shows accuracy for Precion@k for each entity type. We observe high values ($\geq0.80$) during testing Precsion@1 for all entity types, except {\small{\sf City}} (0.73) and {\small{\sf ComicsCharacter}} (0.76). This indicates that our system can return the correct entity linking with very high precision. If we measure the effectiveness of our system at Precsion@5, we observe that our system returns the correct result for each entity, independent of the type, and with a high accuracy of $\geq 0.93$. We report an  accuracy of $\geq 0.95$ at Precision@10 for all entity types with a perfect result for entity type university. Note that the columns \emph{train} denote ``ideal systems'' for the given training data. We observe that even an ideal system fails in rare cases for {\small{\sf Astronaut}}, {\small{\sf WrittenWork}}, {\small{\sf City}}, {\small{\sf Food}} and {\small{\sf Airport}}.

\paragraph{Precision@1 \textgreater 0.6 for cold start entities} Our cold start scenario measures how well our system can detect recently inserted entity matches, in particular, if it has neither seen these entities as relational nor in sentence data. This experiment is particularly challenging for IDEL. During training the system could not observe any contextual clues, neither from positive nor from negative examples. Hence, we test if the system can detect in  such a challenging scenario entities without re-learning embeddings and similarity functions. Table \ref{tab:accuracy} reports measures for such cold-start-entities in the columns \emph{unseen}. We observe that even for Precsion@1, it still achieves decent measures (i.e.\textgreater 0.6) for all entity types, except {\small{\sf Airport}} and {\small{\sf ComicsCharacter}}.


\paragraph{Execution Engine}
In certain practical scenarios it is crucial to be able to execute entity linking in minutes to hours. Consider the brand monitoring example that triggers hourly alerts to a business department about new products or statements in blogs about new products. Typically, business demands here to react within a few hours to place new advertisements or to inform customers.  Therefore, it is crucial that IDEL executes entity linking in minutes to hours.  Moreover, we already discussed that IDEL can recognize potentially unseen entities with a decent Precision@1\textgreater0.6. Ideally, IDEL should regularly update its embeddings and similarity functions asynchronously in the background so as to raise Precision@1 to 0.85 as reported in our experiments in testing.

Table \ref{tab:runtime} reports execution times averaged over all entity types. We observe  for steps at data loading time, such as embed sentences and relational tuples, an average of 208 seconds. For the query execution time
and creating candidate tuples, storing embeddings, applying the similarity metric, ranking and pruning TopK entity mappings, we observe an average of 116 seconds.
Our conclusion is that once a user has set up in IDEL an initial query mapping from entity types in relational data to sentences in text data, the system can asynchronous rebuild embeddings in the background to achieve very high Precision@1 values even for unseen entities for the next time the user hits the same query.

\begin{table}
	\centering
	\begin{tabular}{|l|l|r|}
		\hline
		Phase & Step & RT (sec)                                    \\                               \hline
		Load Time & Loading Model             & 30.50           \\ \hline
		
		Load Time & UDF:EmbedTuples           & 55.00            \\ \hline
		Load Time & UDF:EmbedSentences        & 150.00           \\ \hline
		Load Time & Create Index for Tuples & 0.04 \\ \hline
		Load Time & Create Index for Sentences & 3.07 \\ \hline
		\hline
		Load Time & Sum over all steps                      & 208.1       \\ \hline
		\hline
		Query Time & Cross Join Top10      			& 115.9   \\ \hline
		Query Time & UDF:QueryNN Top10 Sent. & 9.60     \\ \hline
		Query Time & UDF:QueryNN Top10 Tuples 	& 29.15     \\ \hline
		\hline
	\end{tabular}
	\caption{Runtime (in seconds) of different stages in IDEL for the entity type Building. }
	\label{tab:runtime}
\end{table}

\subsection{Error Analysis and Discussion}
\paragraph{Understanding sampling  and computing similarity function} To understand the behavior of IDEL we conducted a closer inspection on results and individual components. Figure~\ref{fig:batches} shows six snapshots from the joint embedding space in IDEL during the training of the similarity function. For example, Figure~\ref{fig:batches}(a)  visualizes on the right a cluster of  58 different entities of the type \emph{building} in 380 tuples, while the left cluster denotes 2377 sentences mentioning these entities. Colors indicate distinct entities\footnote{To keep the colors in the figures somewhat distinguishable, we show here only the most frequent entities, instead of all 58 of them.}.
Figure~\ref{fig:batches}(b)..(f) show in steps of 100 batches how during training new instances (see Section~\ref{par:sampling}), the shape of these clusters change.
Finally, Figure~\ref{fig:batches}(f) shows clusters which combine sentence and relational representations for the same entity. However, we also observe ``yellow'' and ``red'' entities with fewer training examples compared to the ``blue'' and ``light blue'' entities. Our explanation is that the contrastive pairwise loss function does not have sufficient ``signals'' gained from training samples yet to cluster these entities as well.

\paragraph{Performance for unseen entities suffers from sparse attribute density or too few sentences}
IDEL can recognize unseen data with a decent Precision@1\textgreater0.6, except for {\small{\sf Airport}} and {\small{\sf ComicsCharacter}}. This performance is comparable with other state-of-the-art entity linking systems (see \cite{ji2016overview}).  The low performance for the type {\small{\sf Airport}} is most probably due to the extreme sparse average tuple density. As a result, during training the model often retrieves relational tuples with low information gain and many NULL-values. A closer inspection reveals that several errors for this type are disambiguation errors for potential homonyms and undiscovered synonyms. The type {\small{\sf ComicsCharacter}} also performs poorly for unseen entities  compared to other types. This type has the second lowest ratio for Sentence/Instance. Hence, each distinct comic character is represented on average by 18 sentences. The popularity of text data, such as  comic characters, often follows a Zipf distribution. In fact, we inspected our set of comic characters and observed that a few characters are described by the majority of sentences, while most characters are described by only a few sentences. As a result, the system could not learn enough variances to distinguish among these seldom mentioned characters.

\section{Related Work}
\label{sec:related}
Currently, the computational linguistics, the web and the database communities work independently on important aspects of IDEL. However, we are not aware of any system that provides the combined functionality of IDEL. Below we discuss existing work in the areas of text databases, embeddings in databases and entity linking.

\paragraph{Text Databases} Authors of Deep Dive~\citep{shin2015incremental}, InstaRead~\citep{hoffmann2015extreme} and System-T~\citep{chiticariu2010systemt}  propose declarative SQL-based query languages for integrating relational data with text data. Those RDBMS extensions leverage built-in query optimization, indexing and security techniques. They rely on explicitly modeled features for representing syntactic, semantic and lexical properties of an entity in the relational model.
In this work we extend the text database system INDREX~\citep{kilias2015indrex, schneiderinteractive} with the novel functionality of linking relational data to text data. Thereby, we introduce neural embeddings in a  main memory database system, effectively eliminating the need for explicit feature modeling. Our execution system for INDREX is MonetDB, a well known main memory database system~\cite{DBLP:journals/cacm/BonczKM08}. To our best knowledge, no other database system so far provides this functionality for text data.

\paragraph{Embeddings in Databases} Only since recently, authors of~\cite{Bordawekar:2017:UWE:3076246.3076251} investigated methods for integrating vector space embeddings for neural network in relational databases and query processing.   Authors focus on latent information in text data and other  types of data, e.g. numerical values, images and dates. For these data types they embed the latent information for each row in the same table with word2vec~\cite{DBLP:conf/nips/MikolovSCCD13} in the same vector space. Finally, they run queries against this representation for retrieving similar rows. Similar to our work, they suggest to compute embeddings for each row and to access embeddings via UDFs. Our approach goes much further, since we embed latent information from at least two tables in the same vector space, one representing entities and attributes while the other representing spans of text data. Because of the nature of the problem, we can not assume that both representations provide similar characteristics in this vector space. Rather, we need to adopt complex techniques such as SkipThought and pair-wise loss functions to compute similarity measures.

Recently, the information retrieval community recognized the importance of end-to-end information retrieval with neural networks (see~\cite{DBLP:conf/wsdm/MitraC17} for a tutorial). Authors suggest to encode attributes in an indexed table as embeddings to answer topical queries against them. Again, our work goes significantly beyond their ideas and integrates information  across text and relational data inside a database.

\paragraph{Entity Linking and knowledge base completion}
Entity linking is a well-researched problem in computational linguistics\footnote{See \url{http://nlp.cs.rpi.edu/kbp/2017/elreading.html}}. Recently, embeddings have been proposed to jointly represent entities in text and  knowledge graphs~\cite{wang_knowledge_2014}.  Authors of~\cite{lin2015learning} use an embedding for relations and entities in the triple format based on the structures of graphs. However, they do not incorporate additional attributes for the entities into the embedding; also, they only learn an embedding for binary relations, not for n-ary relations. At a very high level, we also apply similar techniques for representing entities in embeddings. However, our approach is based on SkipThought and a pair wise loss function  which works particularly well with many classes (each entity represents its own class) and for sparse data, two data characteristics often found in practical setups for relational databases. Moreover,  our approach is not restricted to triple-based knowledge bases. We can learn an embedding for arbitrary n-ary relations and incorporate their attributes and related entities. Finally, we are not aware of any work that incorporates  neural network based knowledge representation methods  into the query processor of an RDBMS.
%
%

Authors of~\citep{moro2014entity} assign each entity a set of potentially related entities and additional words from sentences mentioning the entity. They weight and prune this signature in a graph, and extract, score and assign subgraphs as semantic signature for each entity. In our work, the idea of a signature is captured by describing an entity via the relation which includes a primary key for the entity and the depending foreign key relations.
Further, our work is orthogonal to ~\citep{moro2014entity}; we represent entity information in the vector space and with neural embeddings and execute the system in a database.

%

\section{Summary and Outlook}
\label{sec:conclusion}
IDEL combines in a single system relational and text representations of entities and capabilities for entity linking. The ability to define Python routines in MonetDB permits us for the first time to conduct this task in a single system, with zero data shipping cost and negligible data transformation costs. To execute this powerful functionality, we have extended MonetDB with UDFs to execute neural embeddings to represent such entities in joint embedding spaces, to compute similarities  based on the idea of the pair-wise contrastive loss and to speed up candidate retrieval with nearest neighbor indexing structures. Therefore, the novelty of our work is in the representation of text data and relational data in the same space, the classification method for entity linking and in a single integrated architecture.

To our best knowledge, this is the first working database system which permits executing such queries on neural embeddings in an RDBMS. As a result, organizations will be able to obtain licenses for a single system only, do not need to hire additional trained linguists, and avoid costly data integration efforts between multiple systems, such as the RDBMS, the text data system (e.g. Lucene) and a homegrown entity linking system. Finally, organizations can reuse trusted and existing efforts from RDBMS, such as security, user management and query optimization techniques.

In our future work we plan to investigate potentially more complex neural architectures, because they are likely able to adapt better to the text and relational data. For example, we are currently limited by the vocabulary of the pre-trained SkipThought. We also plan to use a hybrid model that considers large external linguistic corpora  (as we currently do) and in addition very specific, potentially domain focused corpora from the text database to create improved text embeddings or even character embeddings. Finally, we will investigate deeper effects of other distance functions, such as the \emph{word mover's distance} \cite{DBLP:conf/icml/KusnerSKW15}.




\section*{Acknowledgments}
Our work is funded by the German Federal Ministry of Economic Affairs and Energy (BMWi) under grant agreement 01MD16011E (Medical Allround-Care Service Solutions), grant agreement \\01MD15010B (Smart Data Web) and by the European Union\lq s Horizon 2020 research and innovation program under grant agreement No 732328 (FashionBrain).

\bibliographystyle{abbrv}
\bibliography{DATEXIS,newbibtex}

\begin{thebibliography}{10}

\bibitem{DBLP:conf/dl/AgichteinG00}
E.~Agichtein and L.~Gravano.
\newblock \emph{Snowball}: extracting relations from large plain-text
  collections.
\newblock In {\em {ACM} {DL}}, pages 85--94, 2000.

\bibitem{DBLP:conf/coling/ArnoldDL16}
S.~Arnold, R.~Dziuba, and A.~L{\"o}ser.
\newblock {{TASTY}}: {{Interactive Entity Linking As}}-{{You}}-{{Type}}.
\newblock In {\em {{COLING}}'16 {{Demos}}}, pages 111--115, 2016.
\newblock 00000.

\bibitem{DBLP:conf/sisap/AumullerBF17}
M.~Aum{\"{u}}ller, E.~Bernhardsson, and A.~Faithfull.
\newblock Ann-benchmarks: {A} benchmarking tool for approximate nearest
  neighbor algorithms.
\newblock In {\em {SISAP} 2017}.

\bibitem{Bishop:2006:PRM:1162264}
C.~M. Bishop.
\newblock {\em Pattern Recognition and Machine Learning (Information Science
  and Statistics)}.
\newblock Springer-Verlag New York, Inc., 2006.

\bibitem{DBLP:journals/cacm/BonczKM08}
P.~A. Boncz, M.~L. Kersten, and S.~Manegold.
\newblock {Breaking the memory wall in MonetDB}.
\newblock {\em Commun. {ACM}}, 51(12):77--85, 2008.

\bibitem{Bordawekar:2017:UWE:3076246.3076251}
R.~Bordawekar and O.~Shmueli.
\newblock Using word embedding to enable semantic queries in relational
  databases.
\newblock DEEM'17, pages 5:1--5:4. ACM.

\bibitem{DBLP:conf/stoc/Charikar02}
M.~Charikar.
\newblock Similarity estimation techniques from rounding algorithms.
\newblock In {\em STOC 2002}.

\bibitem{chiticariu2010systemt}
L.~Chiticariu, R.~Krishnamurthy, Y.~Li, S.~Raghavan, F.~R. Reiss, and
  S.~Vaithyanathan.
\newblock {{SystemT}}: An algebraic approach to declarative information
  extraction.
\newblock ACL '10.

\bibitem{DBLP:conf/emnlp/GuptaSR17}
N.~Gupta, S.~Singh, and D.~Roth.
\newblock Entity linking via joint encoding of types, descriptions, and
  context.
\newblock In {\em {EMNLP} 2017}.

\bibitem{tac2016}
H.~D. S.~H. H~Ji, J~Nothman.
\newblock Overview of tac-kbp2016 tri-lingual edl and its impact on end-to-end
  cold-start kbp.
\newblock In {\em TAC}, 2016.

\bibitem{harris1954distributional}
Z.~S. Harris.
\newblock Distributional {{Structure}}.
\newblock {\em \emph{WORD}}, 10(2-3):146--162, Aug. 1954.

\bibitem{tac2017}
B.~Z. J. N. J. M. P.~M. Heng~Ji, Xiaoman~Pan and C.~Costello.
\newblock Overview of tac-kbp2017 13 languages entity discovery and linking.
\newblock In {\em TAC}, 2017.

\bibitem{hoffmann2015extreme}
R.~Hoffmann, L.~Zettlemoyer, and D.~S. Weld.
\newblock Extreme {{Extraction}}: {{Only One Hour}} per {{Relation}}.
\newblock {\em arXiv:1506.06418}, 2015.

\bibitem{ji2016overview}
H.~Ji, J.~Nothman, H.~T. Dang, and S.~I. Hub.
\newblock Overview of tac-kbp2016 tri-lingual edl and its impact on end-to-end
  cold-start kbp.
\newblock {\em TAC'16}.

\bibitem{kilias2015indrex}
T.~Kilias, A.~L{\"o}ser, and P.~Andritsos.
\newblock {{INDREX}}: {{In}}-{{Database Relation Extraction}}.
\newblock {\em Information Systems}, 53:124--144, 2015.

\bibitem{kiros2014unifying}
R.~Kiros, R.~Salakhutdinov, and R.~S. Zemel.
\newblock Unifying {{Visual}}-{{Semantic Embeddings}} with {{Multimodal Neural
  Language Models}}.
\newblock {\em arXiv:1411.2539 [cs]}, Nov. 2014.

\bibitem{kiros2015skip}
R.~Kiros, Y.~Zhu, R.~R. Salakhutdinov, R.~Zemel, R.~Urtasun, A.~Torralba, and
  S.~Fidler.
\newblock Skip-thought vectors.
\newblock In {\em NIPS'15}.

\bibitem{DBLP:conf/icml/KusnerSKW15}
M.~J. Kusner, Y.~Sun, N.~I. Kolkin, and K.~Q. Weinberger.
\newblock From word embeddings to document distances.
\newblock In {\em {ICML}'15}.

\bibitem{lin2015learning}
Y.~Lin, Z.~Liu, M.~Sun, Y.~Liu, and X.~Zhu.
\newblock Learning entity and relation embeddings for knowledge graph
  completion.
\newblock In {\em Proceedings of {{AAAI}}}, 2015.

\bibitem{mikolov2013distributed}
T.~Mikolov, I.~Sutskever, K.~Chen, G.~S. Corrado, and J.~Dean.
\newblock Distributed {{Representations}} of {{Words}} and {{Phrases}} and
  their {{Compositionality}}.
\newblock In {\em {{NIPS}}'13}.

\bibitem{DBLP:conf/nips/MikolovSCCD13}
T.~Mikolov, I.~Sutskever, K.~Chen, G.~S. Corrado, and J.~Dean.
\newblock Distributed representations of words and phrases and their
  compositionality.
\newblock In {\em {NIPS}'13}.

\bibitem{DBLP:conf/wsdm/MitraC17}
B.~Mitra and N.~Craswell.
\newblock Neural text embeddings for information retrieval.
\newblock In {\em {WSDM}'17}.

\bibitem{moro2014entity}
A.~Moro, A.~Raganato, and R.~Navigli.
\newblock Entity {{Linking}} meets {{Word Sense Disambiguation}}: {{A Unified
  Approach}}.
\newblock In {\em {{TACL}}'14}.

\bibitem{DBLP:journals/corr/Perez-Beltrachini17}
L.~Perez{-}Beltrachini and C.~Gardent.
\newblock Analysing data-to-text generation benchmarks.
\newblock {\em CoRR}, abs/1705.03802, 2017.

\bibitem{DBLP:conf/coling/Perez-Beltrachini16}
L.~Perez{-}Beltrachini, R.~Sayed, and C.~Gardent.
\newblock Building {RDF} content for data-to-text generation.
\newblock In {\em {COLING} 2016}.

\bibitem{Raasveldt:2016:VUC:2949689.2949703}
M.~Raasveldt and H.~M\"{u}hleisen.
\newblock {Vectorized UDFs in Column-Stores}.
\newblock SSDBM '16. ACM.

\bibitem{DBLP:conf/sigmod/RenEJH16}
X.~Ren, A.~El{-}Kishky, H.~Ji, and J.~Han.
\newblock Automatic entity recognition and typing in massive text data.
\newblock In {\em {SIGMOD}'16}.

\bibitem{schneiderinteractive}
R.~Schneider, C.~Guder, T.~Kilias, A.~L{\"o}ser, J.~Graupmann, and O.~Kozachuk.
\newblock Interactive {{Relation Extraction}} in {{Main Memory Database
  Systems}}.
\newblock {\em {COLING}'16}.

\bibitem{shin2015incremental}
J.~Shin, S.~Wu, C.~Zhang, F.~Wang, and C.~R{\'e}.
\newblock Incremental {{Knowledge Base Construction Using DeepDive}}.
\newblock {\em VLDB 2015}.

\bibitem{wang_knowledge_2014}
Z.~Wang, J.~Zhang, J.~Feng, and Z.~Chen.
\newblock Knowledge graph embedding by translating on hyperplanes.
\newblock In {\em {AAAI}'14}.

\end{thebibliography}

\end{document}